%% file: main.tex
\documentclass{vldb}

\usepackage{balance}
\usepackage[utf8]{inputenc}
\usepackage{totcount}
\usepackage{url}
\usepackage{libertine}
\usepackage{listings}
\usepackage[T1]{fontenc}
\usepackage{xcolor}
\usepackage{quoting}
\usepackage{multirow}
\usepackage{rotating}
\usepackage{enumitem}
\usepackage{algorithm}
\usepackage[noend]{algpseudocode}
\algrenewcommand\algorithmicrequire{\textbf{Input:}}
\algrenewcommand\algorithmicensure{\textbf{Output:}}

\vldbTitle{Cloudburst: Stateful Functions-as-a-Service}
\vldbAuthors{Vikram Sreekanti, Chenggang Wu, Xiayue Charles Lin, Johann Schleier-Smith, Jose M. Faleiro, Joseph E. Gonzalez, Joseph M. Hellerstein, Alexey Tumanov}

\vldbDOI{https://doi.org/10.14778/3407790.3407836}
\vldbVolume{13}
\vldbNumber{11}
\vldbYear{2020}

\setitemize{noitemsep,topsep=0pt,parsep=0pt,partopsep=0pt,leftmargin=9pt,itemindent=9pt}
\setenumerate{noitemsep,topsep=0pt,parsep=0pt,partopsep=0pt,leftmargin=9pt,itemindent=9pt,labelsep=0.25em}

\input{miscellanea.tex}

\begin{document}


\title{Cloudburst: Stateful Functions-as-a-Service}

\author{
\alignauthor{
    Vikram Sreekanti, Chenggang Wu, Xiayue Charles Lin, Johann Schleier-Smith, \\ Joseph E. Gonzalez, Joseph M. Hellerstein, Alexey Tumanov$^\dagger$ \\
    \affaddr{U.C. Berkeley, $^\dagger$Georgia Tech} \\
    \email{\{vikrams, cgwu, charles.lin, jssmith, jegonzal, hellerstein\}@berkeley.edu, atumanov3@gatech.edu}
}
}


\maketitle

\input{\texdir/abstract.tex}
    
\input{\texdir/intro-jmh.tex}
\input{\texdir/motivation.tex}
\input{\texdir/programming.tex}

\input{\texdir/architecture.tex}
\input{\texdir/consistency.tex}

\input{\texdir/eval.tex}
\input{\texdir/related.tex}

\input{\texdir/conclusion.tex}

\newpage
\bibliographystyle{abbrv}
\bibliography{references}

\end{document}

%% file: miscellanea.tex
\newtoggle{shownotes}
\toggletrue{shownotes}
\newtotcounter{numnotes}

\definecolor{darkgreen}{RGB}{0, 128, 0}
\definecolor{background}{rgb}{0.94,0.95,0.96}

\newcommand{\cuttext}[1]{}

\newcommand{\system}{Cloudburst}
\newcommand{\fullsystem}{\system{}}
\newcommand{\sand}{\texttt{SAND}}

\newcommand{\figwidth}{.46\textwidth}

\newtoggle{showtodos}
\toggletrue{showtodos}
\newtotcounter{numtodos}

\newcommand{\smallitem}[1]{\vspace{0.3em}\noindent\textbf{#1}}
\newcommand{\smallitembot}{\vspace{0.5em}\noindent}

\definecolor{deepblue}{rgb}{0,0,0.5}
\definecolor{deepred}{rgb}{0.6,0,0}
\definecolor{deepgreen}{rgb}{0,0.5,0}

\newcommand\pythonstyle{\lstset{
language=Python,
otherkeywords={self, def, CloudburstClient, CloudburstReference},             
keywordstyle=\color{deepblue},
backgroundcolor=\color{background},
emph={CloudburstClient},          
emphstyle=\color{deepred},    
stringstyle=\color{deepgreen},
showstringspaces=false,            %
basicstyle=\linespread{1.0}\scriptsize\ttfamily,
breaklines=true,
numbers=left,
numberstyle=\ttfamily,
columns=fixed,
basewidth=0.52em,
xleftmargin=6mm,
xrightmargin=0mm,
numberblanklines=true,
}}

\let\origthelstnumber\thelstnumber
\makeatletter
\newcommand*\Suppressnumber{%
  \lst@AddToHook{OnNewLine}{%
    \let\thelstnumber\relax%
     \advance\c@lstnumber-\@ne\relax%
    }%
}

\newcommand*\Reactivatenumber[1]{%
  \setcounter{lstnumber}{\numexpr#1-1\relax}
  \lst@AddToHook{OnNewLine}{%
   \let\thelstnumber\origthelstnumber%
   \refstepcounter{lstnumber}
  }%
}

\lstnewenvironment{python}[1][]
{
\pythonstyle
\lstset{#1}
}
{}

%% file: tex/abstract.tex
\begin{abstract}

Function-as-a-Service (FaaS) platforms and ``serverless'' cloud computing are becoming increasingly popular due to ease-of-use and operational simplicity.
Current FaaS offerings are targeted at stateless functions that do minimal I/O and communication. 
We argue that the benefits of serverless computing can be extended to a broader range of applications and algorithms while maintaining the key benefits of existing FaaS offerings.
We present the design and implementation of \system{}, a stateful FaaS platform that provides familiar Python programming with low-latency mutable state and communication, while maintaining the autoscaling benefits of serverless computing. 
\system{} accomplishes this by leveraging Anna, an autoscaling key-value store, for state sharing and overlay routing combined with mutable caches co-located with function executors for data locality.
Performant cache consistency emerges as a key challenge in this architecture. 
To this end, \system{} provides a combination of \emph{lattice-encapsulated} state and new definitions and protocols for \textit{distributed session consistency}. 
Empirical results on benchmarks and diverse applications show that \system{} makes stateful functions practical, reducing the state-management overheads of current FaaS platforms by orders of magnitude while also improving the state of the art in serverless consistency.

\end{abstract}

%% file: tex/intro-jmh.tex
\section{Introduction} \label{sec:intro}

Serverless computing has become increasingly popular in recent years, with a focus on autoscaling Function-as-a-Service (FaaS) systems.
FaaS platforms allow developers to write functions in standard languages and deploy their code to the cloud with reduced administrative burden. 
The platform is responsible for transparently autoscaling resources from zero to peak load and back in response to workload shifts. 
Consumption-based pricing ensures that developers' cost is proportional to usage of their code: there is no need to overprovision to match peak load, and there are no compute costs during idle periods.
These benefits have made FaaS platforms an attractive target for both research~\cite{excamera, pywren, akkus2018sand, bv-serverless, serverless-cidr19, pocket, baldini2017serverless, van2017spec, gan2019open, fouladi2019laptop} and industry applications~\cite{awscasestudies}.

The hallmark autoscaling feature of serverless platforms is enabled by an increasingly popular design principle: the disaggregation of storage and compute services~\cite{han2013network}. 
Disaggregation allows the compute layer to quickly adapt computational resource allocation to shifting workload requirements, packing functions into VMs while reducing data movement.
Similarly, object or key-value stores can pack multiple users' data storage and access workloads into shared resources with high volume and often at low cost.
Disaggregation also enables allocation at multiple timescales: long-term storage can be allocated separately from short-term compute leases.
Together, these advantages enable efficient autoscaling. 
User code consumes expensive compute resources as needed and accrues only storage costs during idle periods. 

\begin{figure}[t]
  \centering
    \includegraphics[width=\figwidth]{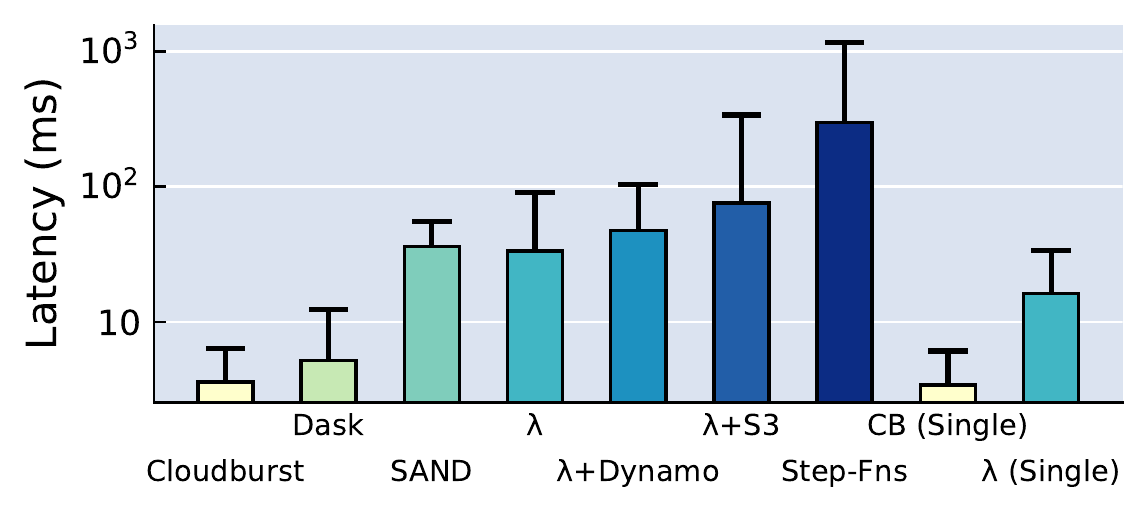}
  \caption{
      Median (bar) and 99th percentile (whisker) latency for \texttt{square(increment(x: int))}.
      \system{} matches the best distributed Python systems and outperforms other FaaS systems by over an order of magnitude ($\S$\ref{sec:eval-microbench}).
  }
  \label{fig:e2e-latency}
  \vspace{-0.5em}
\end{figure}

Unfortunately, today's FaaS platforms take disaggregation to an extreme, imposing significant constraints on developers.
First, the autoscaling storage services provided by cloud vendors---e.g., AWS S3 and DynamoDB---are too high-latency to access with any frequency~\cite{wang2018peeking,serverless-cidr19}.
Second, function invocations are isolated from each other: FaaS systems disable point-to-point network communication between functions. 
Finally, and perhaps most surprisingly, current FaaS offerings have very slow nested function calls: argument- and result-passing is a form of cross-function communication and exhibits the high latency of current serverless offerings~\cite{akkus2018sand}. 
We return these points in $\S$\ref{sec:motiv-pain}, but in short, today's popular FaaS platforms only work well for isolated, \emph{stateless} functions.

As a workaround, many applications---even some that were explicitly designed for serverless platforms---are forced to step outside the bounds of the serverless paradigm altogether.
For example, the ExCamera serverless video encoding system~\cite{excamera} depends upon a single server machine as a coordinator and task assignment service.
Similarly, numpywren~\cite{numpywren} enables serverless linear algebra but provisions a static Redis machine for low-latency access to shared state for coordination.
These workarounds might be tenable at small scales, but they architecturally reintroduce the scaling, fault tolerance, and management problems of traditional server deployments.

\subsection{Toward Stateful Serverless via LDPC}

Given the simplicity and economic appeal of FaaS, we are interested in exploring designs that preserve the autoscaling and operational benefits of current offerings, while adding performant, cost-efficient, and consistent shared state and communication.
This ``stateful'' serverless model opens up autoscaling FaaS to a much broader array of applications and algorithms.
We aim to demonstrate that serverless architectures can support stateful applications while maintaining the simplicity and appeal of the serverless programming model.

For example, many low-latency services need to autoscale to handle bursts while dynamically manipulating data based on request parameters. 
This includes webservers managing user sessions, discussion forums managing threads, ad servers managing ML models, and more.
Similarly, many parallel and distributed protocols require fine-grained messaging, from quantitative tasks like distributed aggregation~\cite{kempe2003gossip} to system tasks like membership~\cite{das2002swim} or leader election~\cite{awerbuch1987optimal}.
These protocols forms the backbone of parallel and distributed systems. 
As we see in $\S$\ref{sec:eval}, these scenarios are infeasible on today's stateless FaaS platforms.

To enable stateful serverless computing, we propose a new design principle: \emph{logical disaggregation with physical colocation} (LDPC). 
Disaggregation is needed to provision, scale, and bill storage and compute independently, but we want to deploy resources to different services in close physical proximity.
In particular, a running function's ``hot'' data should be kept physically nearby for low-latency access.
Updates should be allowed at any function invocation site, and cross-function communication should work at wire speed.

Colocation of compute and data is a well-known method to overcome performance barriers, but it can raise thorny correctness challenges at the compute layer.
For locality, each compute node should be able to independently update its copy of the data. 
However, given that a single composition of multiple functions may run across multiple nodes, we need to preserve a consistent ``session''~\cite{terry1994session} that is distributed across those nodes.
\emph{Distributed session consistency} is a challenge we must address in this context.

\subsection{\system{}: A Stateful FaaS Platform} \label{sec:intro-star}

In this paper, we present a new Function-as-a-Service platform called \fullsystem{} that removes the shortcomings of commercial systems highlighted above, without sacrificing their benefits. 
\system{} is unique in achieving logical disaggregation and physical colocation of computation and state, and in allowing programs written in a traditional language to observe consistent state across function compositions. 
\system{} is designed to be an autoscaling Functions-as-a-Service system---similar to AWS Lambda or Google Cloud Functions---but with new abstractions that enable performant, stateful programs.

\system{} achieves this via a combination of an autoscaling key-value store (providing state sharing and overlay routing) and mutable caches co-located with function executors (providing data locality).
The system is built on top of Anna~\cite{wu2019anna, anna-vldb}, a low-latency autoscaling key-value store designed to achieve a variety of coordination-free consistency levels by using mergeable monotonic \emph{lattice} data structures~\cite{shapiro2011conflict, Conway:2012:LLD:2391229.2391230}.
For performant consistency, \system{} takes advantage of Anna's design by transparently encapsulating opaque user state in lattices so that Anna can consistently merge concurrent updates.
In addition, we present novel protocols that ensure consistency guarantees (repeatable read and causal consistency) across function invocations that run on separate nodes.
We evaluate \system{} via microbenchmarks as well as two application scenarios using third-party code, demonstrating benefits in performance, predictable latency, and consistency. 
In sum, this paper's contributions include:

\begin{enumerate}\itemsep0em
    \item The design and implementation of an autoscaling serverless architecture that combines logical disaggregation with physical co-location of compute and storage (LDPC)~($\S$\ref{sec:arch}).

    \item  Identification of \emph{distributed session consistency} concerns and new protocols to achieve two distinct distributed session consistency guarantees---repeatable read and causal consistency---for compositions of functions ($\S$\ref{sec:consistency}).
    
    \item The ability for programs written in traditional languages to enjoy coordination-free storage consistency for their native data types via \emph{lattice capsules} that wrap program state with metadata that enables automatic conflict  APIs supported by Anna~($\S$\ref{sec:consistency-lattice}). 

    \item An evaluation of \system{}'s performance and consistency on workloads involving state manipulation, fine-grained communication and dynamic autoscaling~($\S$\ref{sec:eval}).
\end{enumerate}



%% file: tex/motivation.tex
\section{Motivation and Background}\label{sec:motivation}

\begin{figure}[t]
\begin{python}
from cloudburst import *
cloud = CloudburstClient(cloudburst_addr, my_ip)
cloud.put('key', 2)
reference = CloudburstReference('key')
def sqfun(x): return x * x
sq = cloud.register(sqfun, name='square')

print('result: 
> result: 4 

future = sq(3, store_in_kvs=True) 
print('result: 
> result: 9
\end{python}
\caption{A script to create and execute a \system{} function.} 
\label{lst:simple-func}
\vspace{-0.1em}
\end{figure}

Although serverless infrastructure has gained traction recently, there remains significant room for improvement in performance and state management.
In this section, we discuss common pain points in building applications on today's serverless infrastructure ($\S$\ref{sec:motiv-pain}) and explain \system{}'s design goals ($\S$\ref{sec:motiv-challenges}).

\subsection{Deploying Serverless Functions Today} \label{sec:motiv-pain}

Current FaaS offerings are poorly suited to managing shared state, making it difficult to build applications, particularly latency-sensitive ones.
There are three kinds of shared state management that we focus on in this paper: function composition, direct communication, and shared mutable storage.

\smallitem{Function Composition}. 
For developers to embrace serverless as a general programming and runtime environment, it is necessary that function composition work as expected.
Pure functions share state by passing arguments and return values to each other. 
Figure~\ref{fig:e2e-latency} (discussed in $\S$\ref{sec:eval-microbench}), shows the performance of a simple composition of side-effect-free arithmetic functions.
AWS Lambda imposes a latency overhead of up to 20ms for a single function invocation, and this overhead compounds when composing functions. 
AWS Step Functions, which automatically chains together sequences of operations, imposes an even higher penalty.
Since the overheads compound linearly, the overhead of a call stack as shallow as 5 functions saturates tolerable limits for an interactive service ($\sim$100ms). 
Functional programming patterns for state sharing are not an option in current FaaS platforms.

\smallitem{Direct Communication}.
FaaS offerings disable inbound network connections, requiring functions to communicate through high-latency storage services like S3 or DynamoDB.
While point-to-point communication may seem tricky in a system with dynamic membership, distributed hashtables (DHTs) or lightweight key-value stores (KVSs) can provide a lower-latency solution than deep storage for routing messages between migratory function instances~\cite{Ratnasamy:2001:SCN:964723.383072,stoica2001chord,rowstron2001pastry,rodruigues2003one}. 
Current FaaS vendors do not offer autoscaling, low-latency DHTs or KVSs.
Instead, as discussed in $\S$\ref{sec:intro}, FaaS applications resort to server-based solutions for lower-latency storage, like hosted Redis and memcached.

\smallitem{Low-Latency Access to Shared Mutable State}.
Recent studies~\cite{wang2018peeking,serverless-cidr19} have shown that latencies and costs of shared autoscaling storage for FaaS are orders of magnitude worse than underlying infrastructure like shared memory, networking, or server-based shared storage. 
Worse, the available systems offer weak data consistency guarantees.
For example, AWS S3 offers no guarantees across multiple clients (isolation) or for inserts and deletes from a single client (atomicity).
This kind of weak consistency can produce very confusing behavior. 
For example, simple expressions like $f(x, g(x))$ may produce non-deterministic results: $f$ and $g$ are different ``clients'', so there is no guarantee about the versions of $x$ read by $f$ and $g$. 

\subsection{Towards Stateful Serverless} \label{sec:motiv-challenges}

\smallitem{Logical Disaggregation with Physical Colocation}.
As a principle, LDPC leaves significant latitude for designing mechanisms and policy that co-locate compute and data while preserving correctness. 
We observe that many of the performance bottlenecks described above can be addressed by an architecture with distributed storage and local caching.
A low-latency autoscaling KVS can serve as both global storage and a DHT-like overlay network.
To provide better data locality to functions, a KVS cache can be deployed on every machine that hosts function invocations. 
\system{}'s design includes consistent mutable caches in the compute tier~($\S$\ref{sec:arch}).


\smallitem{Consistency}.
Distributed mutable caches introduce the risk of cache inconsistencies, which can cause significant developer confusion.
%
We could implement strong consistency across caches (e.g., linearizability) via quorum consensus (e.g., Paxos~\cite{lamport2001paxos}). 
This offers appealing semantics but has well-known issues with latency and availability~\cite{brewercap,chandra2007paxos}.
In general, consensus protocols are a poor fit for the internals of a dynamic autoscaling framework: consensus requires fixed membership, and membership (``view'') change involves high-latency agreement protocols (e.g.,~\cite{birman1987exploiting}).
Instead, applications desiring strong consistency can employ a slow-changing consensus service adjacent to the serverless infrastructure.

Coordination-free approaches to consistency are a better fit to the elastic membership of a serverless platform.
Bailis, et al.\cite{Bailis:2013:HAT:2732232.2732237} categorized consistency guarantees that can be achieved without coordination.
We chose the Anna KVS~\cite{wu2019anna} as \system{}'s storage engine because it supports all these guarantees.
Like CvRDTs~\cite{shapiro2011conflict}, Anna uses \emph{lattice} data types for coordination-free consistency. That is, Anna values offer a \texttt{merge} operator that is insensitive to batching, ordering and repetition of requests---\texttt{merge} is associative, commutative and idempotent. Anna uses lattice composition~\cite{Conway:2012:LLD:2391229.2391230} to implement consistency;
Anna's lattices are simple and composable as in Bloom~\cite{Conway:2012:LLD:2391229.2391230}; 
we refer readers to~\cite{wu2019anna} for more details.
Anna also provides autoscaling at the storage layer, responding to workload changes by selectively replicating frequently-accessed data, growing and shrinking the cluster, and moving data between storage tiers (memory and disk) for cost savings~\cite{anna-vldb}.

However, Anna only supports consistency for \emph{individual} clients, each with a fixed IP-port pair.
In \system{}, a request like $f(x, g(x))$ may involve function invocations on separate physical machines and requires consistency across functions---we term this \emph{distributed session consistency}.
In $\S$\ref{sec:consistency}, we provide protocols for various consistency levels.

\smallitem{Programmability}.
We want to provide consistency without imposing undue burden on programmers, but
Anna can only store values that conform to its lattice-based type system.
To address this, \system{} introduces \emph{lattice capsules} ($\S$\ref{sec:consistency-lattice}), which transparently wrap opaque program state in lattices chosen to support \system{}'s consistency protocols.
Users gain the benefits of Anna's conflict resolution and \system{}'s distributed session consistency without having to modify their programs.


\smallitembot

We continue with \system{}'s programmer interface.
We return to \system{}'s design in $\S$\ref{sec:arch} and consistency mechanisms in $\S$\ref{sec:consistency}.

%% file: tex/programming.tex
\section{Programming Interface} \label{sec:programming}

\system{} accepts programs written in vanilla Python\footnote{There is nothing fundamental in our choice of Python---we chose it simply because it is a commonly used high-level language.}.
An example client script to execute a function is shown in Figure~\ref{lst:simple-func}.
\system{} functions act like regular Python functions but trigger remote computation in the cloud.
Results by default are sent directly back to the client (line 8), in which case the client blocks synchronously.
Alternately, results can be stored in the KVS, and the response key is wrapped in a \texttt{CloudburstFuture} object, which retrieves the result when requested (line 11-12).

Function arguments are either regular Python objects (line 11) or KVS references (lines 3-4).
KVS references are transparently retrieved by \system{} at runtime and deserialized before invoking the function.
To improve performance, the runtime attempts to execute a function call with KVS references on a machine that might have the data cached.
We explain how this is accomplished in $\S$\ref{sec:arch-fluent-scheduler}.
Functions can also dynamically retrieve data at runtime using the \system{} communication API described below.

To enable stateful functions, \system{} allows programmers to \texttt{put} and \texttt{get} Python objects via the Anna KVS API.
Object serialization and encapsulation for consistency ($\S$\ref{sec:consistency-lattice}) is handled transparently by the runtime.
In the common case, \texttt{put} and \texttt{get} are fast due to the presence of caches at the function executors.

For repeated execution, \system{} allows users to register arbitrary compositions of functions.
We model function compositions as DAGs in the style of systems like Apache Spark~\cite{zaharia2012resilient}, Dryad~\cite{isard2007dryad}, Apache Airflow~\cite{airflow}, and Tensorflow~\cite{abadi2016tensorflow}.
This model is also similar in spirit to cloud services like AWS Step Functions that automatically chain together functions in existing serverless systems.

Each function in the DAG must be registered with the system (line 4) prior to use in a DAG.
Users specify each function in the DAG and how they are composed---results are automatically passed from one DAG function to the next by the \system{} runtime.
The result of a function with no successor is either stored in the KVS or returned directly to the user, as above.
\system{}'s resource management system ($\S$\ref{sec:arch-fluent-mgmt}) is responsible for scaling the number of replicas of each function up and down.

\begin{table}\scriptsize
\caption{The \texttt{\system{}} object API.
Users can interact with the key value store and send and receive messages.
}
\begin{center}
\begin{tabular}{|p{10em}|p{20em}|}
\hline
\textbf{API Name} & \textbf{Functionality} \\
\hline
\texttt{get(key)} & Retrieve a key from the KVS.  \\
\hline
\texttt{put(key, value)} & Insert or update a key in the KVS. \\
\hline
\texttt{delete(key)} & Delete a key from the KVS. \\
\hline
\texttt{send(recv, msg)} & Send a message to another executor.\\
\hline
\texttt{recv()} & Receive outstanding messages for this function. \\
\hline
\texttt{get\_id()} & Get this function's unique ID \\ 
\hline
\end{tabular}
\end{center}
\label{table:api}
\vspace{-.5em}
\end{table}

\smallitem{\system{} System API}.
\system{} provides developers an interface to system services---
Table~\ref{table:api} provides an overview. 
The API enables KVS interactions via \texttt{get} and \texttt{put}, and it enables message passing between function invocations.
Each function invocation is assigned a unique ID, and functions can advertise this ID to well-known keys in the KVS.
These unique IDs are translated into physical addresses and used to support direct messaging.

Note that this process is a generalization of the process that is used for function composition, where results of one function are passed directly to the next function.
We expose these as separate mechanisms because we aimed to simplify the common case (function composition) by removing the hassle of communicating unique IDs and explicitly sharing results.


In practice this works as follows.
First, one function writes its unique ID to a pre-agreed upon key in storage.
The second function waits for that key to be populated and then retrieves the first thread's ID by reading that key.
Once the second function has the first function's ID, it uses the \texttt{send} API to send a message.
When \texttt{send} is invoked, the executor thread uses a deterministic mapping to convert from the thread's unique ID to an IP-port pair. 
The executor thread opens a TCP connection to that IP-port pair and sends a direct message.
If a TCP connection cannot be established, the message is written to a key in Anna that serves as the receiving thread's ``inbox''.
When \texttt{recv} is invoked by a function, the executor returns any messages that were queued on its local TCP port.
On a \texttt{recv} call, if there are no messages on the local TCP port, the executor will read its inbox in storage to see if there are any messages stored there.
The inbox is also periodically read and cached locally to ensure that messages are delivered correctly.


%% file: tex/architecture.tex
\section{Architecture} \label{sec:arch}

\system{} implements the principle of logical disaggregation with physical colocation (LDPC).
To achieve disaggregation, the \system{} runtime autoscales independently of the Anna KVS.
Colocation is enabled by mutable caches placed in the \system{} runtime for low latency access to KVS objects.

\begin{figure}[t]
  \centering
    \includegraphics[width=0.42\textwidth]{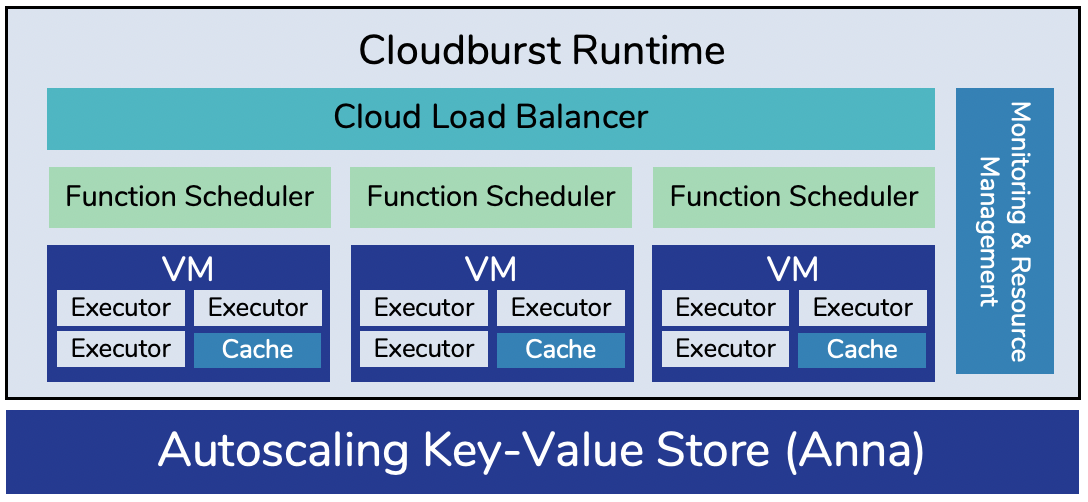}
  \caption{An overview of the \system{} architecture.}
  \label{fig:arch-overview}
\end{figure}

Figure~\ref{fig:arch-overview} provides an overview of the \system{} architecture.
There are four key components: function executors, caches, function schedulers, and a resource management system.
User requests are received by a scheduler, which routes them to function executors.
Each scheduler operates independently, and the system relies on a standard stateless cloud load balancer (AWS Elastic Load Balancer).
Function executors run in individual processes that are packed into VMs along with a local cache per VM.
The cache on each VM intermediates between the local executors and the remote KVS.
All \system{} components are run in individual Docker~\cite{docker} containers.
\system{} uses Kubernetes~\cite{kubernetes} simply to start containers and redeploy them on failure.
\system{} system metadata, as well as persistent application state, is stored in Anna which provides autoscaling and fault tolerance.

\subsection{Function Executors} \label{sec:arch-fluent-executor}

Each \system{} executor is an independent, long-running Python process.
Schedulers ($\S$\ref{sec:arch-fluent-scheduler}) route function invocation requests to executors.
Before each invocation, the executor retrieves and deserializes the requested function and transparently resolves all KVS reference function arguments in parallel.
DAG execution requests span multiple function invocations, and after each DAG function invocation, the runtime triggers downstream DAG functions.
To improve performance for repeated execution ($\S$\ref{sec:programming}), each DAG function is deserialized and cached at one or more executors.
Each executor also publishes local metrics to the KVS, including the executor's cached functions, stats on its recent CPU utilization, and the execution latencies for finished requests.
We explain in the following sections how this metadata is used.

\subsection{Caches} \label{sec:arch-fluent-cache}

To ensure that frequently-used data is locally available, every function execution VM has a local cache, which executors contact via IPC.
Executors interface with the cache, not directly with Anna; the cache issues requests to the KVS as needed.
When a cache receives an update from an executor, it updates the data locally, acknowledges the request, and asynchronously sends the result to the KVS to be merged.
If a cache receives a request for data that it does not have, it asynchronously retrieves it from the KVS.

\system{} must ensure the freshness of data in caches.
A naive (but correct) scheme is for the \system{} caches to poll the KVS for updates, or for the cache to blindly evict data after a timeout.
In a typical workload where reads dominate writes, this generates unnecessary load on the KVS. 
Instead, each cache periodically publishes a snapshot of its cached keys to the KVS.
We modified Anna to accept these cached keysets and incrementally construct an index that maps each key to the caches that store it; Anna uses this index to periodically propagate key updates to caches.
Lattice encapsulation enables Anna to correctly merge conflicting key updates ($\S$\ref{sec:consistency-lattice}).
The index itself is partitioned across storage nodes following the same scheme Anna uses to partition the key space, so Anna takes the index overhead into consideration when making autoscaling decisions.

\subsection{Function Schedulers} \label{sec:arch-fluent-scheduler}

A key goal of \system{}'s architecture is to enable low latency function scheduling.
However, policy design is not a main goal of this paper; \system{}'s scheduling mechanisms allow pluggable policies to be explored in future work.
In this section, we describe \system{}'s scheduling mechanisms, illustrating their use with policy heuristics that enable us to demonstrate
benefits from data locality and load balancing.

\smallitem{Scheduling Mechanisms}.
Schedulers handle requests to register or invoke functions.
New functions are registered by storing them in Anna and updating a shared KVS list of registered functions.
For new DAGs, the scheduler verifies that each function in the DAG exists before picking an executor on which to cache it.

For single function execution requests, the scheduler picks an executor and forwards the request to it.
DAG requests require more work: The scheduler creates a schedule by picking an executor for each DAG function---which is guaranteed to have the function stored locally---and broadcasts this schedule to all participating executors. 
The scheduler then triggers the first function(s) in the DAG and, if the user wants the result stored in the KVS, returns a \texttt{CloudburstFuture}.

DAG topologies are the scheduler's only persistent metadata and are stored in the KVS.
Each scheduler tracks how many calls it receives per DAG and per function and stores these statistics in the KVS.
Finally, each scheduler constructs a local index that tracks the set of keys stored by each cache; this is used for the scheduling policy described next.

\smallitem{Scheduling Policy}.
Our scheduling policy makes heuristic decisions using metadata reported by executors, including cached key sets and executor load.
We prioritize data locality when scheduling both single functions and DAGs.
If the function's arguments have KVS references, the scheduler inspects its local cached key index and picks the executor with the most data cached locally---to take advantage of locality, the user must specify KVS references ($\S$\ref{sec:programming}).
Otherwise, the scheduler picks an executor at random.

Hot data and functions are replicated across many executor nodes via backpressure.
The few nodes initially caching hot keys will quickly be saturated with requests and report high utilization (above 70\%).
The scheduler tracks this utilization and avoids overloaded nodes, picking new nodes instead.
The new nodes will then fetch and cache the hot data, effectively increasing the replication factor and hence the number of options the scheduler has for the next request containing a hot key.

\subsection{Monitoring and Resource Management} \label{sec:arch-fluent-mgmt}

An autoscaling system must track system load and performance metrics to make effective decisions.
\system{} uses Anna as a substrate for metric collection.
Each thread independently tracks an extensible set of metrics (described above) and publishes them to the KVS.
The monitoring system asynchronously aggregates these metrics from storage and uses them for its policy engine.

For each DAG, the monitoring system compares the incoming request rate to the number of requests serviced by executors.
If the incoming request rate is significantly higher than the request completion rate of the system, the monitoring engine will increase the resources allocated to that DAG function by pinning the function onto more executors. 
If the overall CPU utilization of the executors exceeds a threshhold (70\%), then the monitoring system will add nodes to the system.
Similarly, if executor utilization drops below a threshold (20\%), we deallocate resources accordingly.
We rely on Kubernetes to manage our clusters and efficiently scale the cluster.
This simple approach exercises our monitoring mechanisms and provides adequate behavior (see $\S$\ref{sec:eval-scaling}).

When a new node is allocated, it reads the relevant data and metadata (e.g., functions, DAG metadata) from the KVS.
This allows Anna to serve as the source of truth for system metadata and removes concerns about efficiently scaling the system.

The heuristics that we described here are based on the existing dynamics of the system (e.g., node spin up time).
We discuss potential advanced auto-scaling mechanisms and policies as a part of Future Work ($\S$\ref{sec:limitations}), which might draw more heavily on understanding how workloads interact with our infrastructure.

\subsection{Fault Tolerance} \label{sec:arch-ft}

At the storage layer, \system{} relies on Anna's replication scheme for $k$-fault tolerance.
For the compute tier, we adopt the standard approach to fault tolerance taken by many FaaS platforms.
If a machine fails while executing a function, the whole DAG is re-executed after a configurable timeout.
The programmer is responsible for handling side-effects generated by failed programs if they are not idempotent.
In the case of an explicit program error, the error is returned to the client.
This approach should be familiar to users of AWS Lambda and other FaaS platforms, which provides the same guarantees. 
More advanced guarantees are a subject for future work~($\S$\ref{sec:limitations}).

%% file: tex/consistency.tex
\section{Cache Consistency} \label{sec:consistency}

As discussed in Section~\ref{sec:programming}, \system{} developers can register compositions of functions as a DAG.
This also serves as the scope of consistency for the programmer, sometimes called a ``session''~\cite{terry1994session}.
The reads and writes in a session together experience the chosen definition of consistency, even across function boundaries.
The simplest way to achieve this is to run the entire DAG in a single thread and let the KVS provide the desired consistency level.
However, to allow for autoscaling and flexible scheduling, \system{} may choose to run functions within a DAG on different executors---in the extreme case, each function could run on a separate executor.
This introduces the challenge of \emph{distributed session consistency}: Because a DAG may run across many machines, the executors involved in a single DAG must provide consistency across different physical machines.


In the rest of this section, we describe distributed session consistency in \system{}.
We begin by explaining two different guarantees ($\S$\ref{sec:consistency-guarantees}), describe how we encapsulate user-level Python objects to interface with Anna's consistency mechanisms~($\S$\ref{sec:consistency-lattice}), and present protocols for the guarantees ($\S$\ref{sec:consistency-protocols}).

\subsection{Consistency Guarantees} \label{sec:consistency-guarantees}

A wide variety of coordination-free consistency and isolation guarantees have been identified in the literature.
We focus on two guarantees here; variants are presented in $\S$\ref{sec:eval-consistency} to illustrate protocol costs. 
In our discussion, we will denote keys with lowercase letters like $k$; $k_v$ is a version $v$ of key $k$.

We begin with repeatable read (RR) consistency. 
RR is adapted from the transactions literature~\cite{berenson1995critique}, hence it assumes sequences of functions---i.e., linear DAGs. 
Given a read-only expression $f(x, g(x))$, RR guarantees that both $f$ and $g$ read the same version $x_v$. More generally:

\textit{\textbf{Repeatable Read Invariant}: In a linear DAG, when any function reads a key $k$, either it sees the most recent update to $k$ within the DAG, or in the absence of preceding updates it sees the first version $k_v$ read by any function in the DAG}\footnote{Note that RR \emph{isolation} also prevents reads of uncommitted version from other transactions~\cite{berenson1995critique}. Transactional isolation is a topic for future work~($\S$\ref{sec:limitations}).}.

The second guarantee we explore is causal consistency, one of the strongest coordination-free consistency models~\cite{COPS, mahajan11cacTR,Bailis:2013:HAT:2732232.2732237}.
In a nutshell, causal consistency requires reads and writes to respect Lamport's ``happens-before'' relation~\cite{Lamport:1978:TCO:359545.359563}.
One key version $k_i$ \emph{influences} another version $l_j$ if a read of $k_i$ happens before a write of $l_j$; we denote this as $k_i \rightarrow l_j$.
Within a \system{} DAG, this means that if a function reads $l_j$, subsequent functions must not see versions of $k$ that happened before $k_i$: they can only see $k_i$, versions concurrent with $k_i$, or versions newer than $k_i$.
Note that causal consistency does not impose restrictions on key versions read by concurrent functions within a DAG.
For example, if function $f$ calls two functions, $g$ and $h$ (executed in parallel), and both $g$ and $h$ read key $k$, the versions of $k$ read by $g$ and $h$ may diverge.
Prior work introduces a variety of causal building blocks that we extend.
Systems like Anna~\cite{wu2019anna} track causal histories of individual objects but do not track ordering between objects.
Bolt-on causal consistency~\cite{Bailis:2013:BCC:2463676.2465279} developed techniques to achieve multi-key causal snapshots at a single physical location.
\system{} must support multi-key causal consistency that spans \emph{multiple} physical sites. 

\textit{\textbf{Causal Consistency Invariant}: 
Consider a function $f$ in DAG $G$ that reads a version $k_v$ of key $k$.
Let $V$ denote the set of key versions read previously by $f$ or by any of $f$'s ancestors in $G$.
Denote the dependency set for $f$ at this point as $D = \{ d_i | d_i \rightarrow l_j \in V\}$. 
The version $k_v$ that is read by $f$ must satisfy the invariant $k_v \not\rightarrow k_i \in D$. 
That is, $k_v$ is concurrent to or newer than any version of $k$ in the dependency set $D$.
}



\subsection{Lattice Encapsulation} \label{sec:consistency-lattice}

Mutable shared state is a key tenet of \system{}'s design ($\S$\ref{sec:programming}), and we rely on Anna's lattices to resolve conflicts from concurrent updates. 
Typically, Python objects are not lattices, so \system{} transparently encapsulates Python objects in lattices.

By default, \system{} encapsulates each bare program value into an Anna \emph{last writer wins} (LWW) lattice---a composition of an Anna-provided global timestamp and the value. 
The global timestamp is generated by each node in a coordination-free fashion by concatenating the local system clock and the node's unique ID.
Anna merges two LWW versions by keeping the value with the higher timestamp.
This allows \system{} to achieve eventual consistency: All replicas will agree on the LWW value that corresponds to the highest timestamp for the key~\cite{vogels2009eventually}.
It also provides timestamps for the RR protocol below.

In causal consistency mode, \system{} encapsulates each key $k$ in a causal lattice---the composition of an Anna-provided vector clock~\cite{Raynal:1996:LTC:226705.226711} that identifies $k$'s version, a dependency set that tracks key versions that $k$ depends on, and the value.
Each vector clock consists of a set of $\langle id, clock \rangle$ pairs where the $id$ is the unique ID of the function executor thread that updated $k$, and the $clock$ is a monotonically-growing logical clock.
Upon merge, Anna keeps the causal consistency lattice whose vector clock dominates.
Vector clock $vc_1$ dominates $vc_2$ if it is at least equal in all entries and greater in at least one; otherwise, $vc_1$ and $vc_2$ are concurrent.
If two vector clocks are concurrent, Anna merges the keys by: (1) creating a new vector clock with the pairwise maximum of each entry in the two keys' vector clock; and (2) merging the dependency sets and values via set union.
In most cases, an object has only one version. 
However, to de-encapsulate a causally-wrapped object with multiple concurrent versions, \system{} presents the user program with one version chosen via an arbitrary but deterministic tie-breaking scheme.
Regardless of which version is returned, the user program sees a causally consistent history; the cache layer retains the concurrent versions for the consistency protocol described below.
Applications can also choose to retrieve all concurrent versions and resolve updates manually.



\subsection{Distributed Session Protocols} \label{sec:consistency-protocols}

\smallitem{Distributed Session Repeatable Read}.
To achieve repeatable read, the \system{} cache on each node creates ``snapshot'' versions of each locally cached object upon first read, and
the cache stores them for the lifetime of the DAG.
When invoking a downstream function in the DAG, we propagate a list of cache addresses and version timestamps for all snapshotted keys seen so far.

In order to maintain repeatable read consistency, the downstream executor needs to read the same versions of variables as read by the upstream executor.
Algorithm~\ref{alg:rr} shows pseudocode for the process that downstream executors follow to ensure this.
When such an executor in a DAG receives a read request for key $k$, it includes the prior version snapshot metadata in its request to the cache ($R$ in Algorithm~\ref{alg:rr}).
If $k$ has been previously read and the exact version is not stored locally, the cache queries the upstream cache that stores the correct version (line 5 in Algorithm~\ref{alg:rr}).
If the exact version is stored locally (line 7), we return it directly to the user.
Finally, if the key has not been read thus far, we return any available version to the client (line 9). 
If the upstream cache fails, we restart the DAG from scratch.
Finally, the last executor in the DAG (the ``sink'') notifies all upstream caches of DAG completion, allowing version snapshots to be evicted.

\begin{algorithm}[t] 
{
    \caption{Repeatable Read}
    \begin{algorithmic}[1]
    \Require $k$, $R$ \\ // $k$ is the requested key; $R$ is the set of keys previously read by the DAG
        \If{$k \in R$}
            \State{$cache\_version = cache.get\_metadata(k)$}
            \If{$cache\_version == NULL \lor cache\_version\ != R[k].version$}
                \State{return $cache.fetch\_from\_upstream(k)$}
            \Else
                \State{return $cache.get(k)$}
            \EndIf
        \Else
            \State{return $cache.get\_or\_fetch(k)$}
        \EndIf
    \end{algorithmic}
    \label{alg:rr}
}
\end{algorithm}

\smallitem{Distributed Session Causal Consistency}.
To support causal consistency in \system{}, we use causal lattice encapsulation and augment the \system{} cache to be a causally consistent store, implementing the bolt-on causal consistency protocol~\cite{Bailis:2013:BCC:2463676.2465279}.  
The protocol ensures that each cache always holds a ``causal cut'': For every pair of versions $a_i, b_j$ in the cache, $\not\exists a_k: a_i \rightarrow a_k, a_k \rightarrow b_j$.
Storing a causal cut ensures that key versions read by functions executed on \textit{one} node (accessing a single causal cut) satisfy the causal consistency invariant discussed in Section~\ref{sec:consistency-guarantees}.

\begin{figure}[t]
  \centering
    \includegraphics[width=0.37\textwidth]{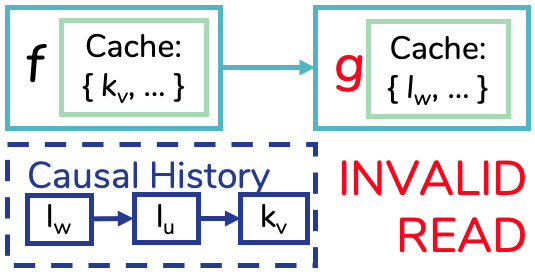}
  \caption{An illustration of a scenario in which two functions executing on separate machines might unwittingly read causally inconsistent data.
  }
  \label{fig:causal-invalid}
\end{figure}

However, maintaining a causal cut within each individual cache is not sufficient to achieve distributed session causal consistency.
Consider a DAG with two functions $f(k)$ and $g(l)$, which are executed in sequence on different machines---Figure~\ref{fig:causal-invalid} illustrates the following situation.
Assume $f$ reads $k_v$ and there is a dependency $l_u \rightarrow k_v$.
If the causal-cut cache of the node executing $g$ is unaware of the constraint on valid versions of $l$, $g$ could read an old version $l_w: l_w \rightarrow l_u$, thereby violating causality.
The following protocol solves this challenge: In addition to shipping read-set metadata (as in RR), each executor ships the set of causal dependencies (pairs of keys and their associated vector clocks) of the read set to downstream executors. 
Caches upstream store version snapshots of these causal dependencies.

\begin{algorithm}[t] 
{
    \caption{Causal Consistency}
    \begin{algorithmic}[1]
    \Require $k$, $R$, $dependencies$ \\ // $k$ is the requested key; $R$ is the set of keys previously read by the DAG; $dependencies$ is the set of causal dependencies of keys in $R$
        \If{$k \in R$}
            \State{$cache\_version = cache.get\_metadata(k)$} \\
\ \ \ \ \ \ // $valid$ returns true if $k \geq cache\_version$
            \If{$valid(cache\_version, R[k])$} 
                \State{return $cache.get(k)$}
            \Else
                \State{return $cache.fetch\_from\_upstream(k)$}
            \EndIf
        \EndIf
        
        \If{$k \in dependencies$}
            \State{$cache\_version = cache.get\_metadata(k)$}
            \If{$valid(cache\_version, dependencies[k])$}
                \State{return $cache.get(k)$}
            \Else
                \State{return $cache.fetch\_from\_upstream(k)$}
            \EndIf
        \EndIf
    \end{algorithmic}
    \label{alg:cc}
}
\end{algorithm}

For each key $k$ requested, the downstream cache first checks whether the locally-cached key's vector clock is causally concurrent with or dominates that of the version snapshot stored at the upstream cache (lines 2-3 of Algorithm~\ref{alg:cc}).
If so, the cache returns the local version (line 5); otherwise, it queries the upstream cache for the correct version snapshot (line 7).
We perform the same check if the key being requested is in the set of dependencies shipped from the upstream cache---if we have a valid version locally, we return it, and otherwise, we fetch a snapshot from the upstream cache (lines 8-13 of Algorithm~\ref{alg:cc}).
This protocol guarantees that key versions read by functions executed across \textit{different} nodes (i.e., different causal cuts) follow the causal consistency invariant, guaranteeing distributed session causal consistency.

%% file: tex/eval.tex
\section{Evaluation} \label{sec:eval}

We now present a detailed evaluation of \system{}.
We first study the individual mechanisms implemented in \system{} ($\S$\ref{sec:eval-microbench}), demonstrating orders of magnitude improvement in latency relative to existing serverless infrastructure for a variety of tasks.
Next we study the overheads introduced by \system{}'s consistency mechanisms ($\S$\ref{sec:eval-consistency}), and finally we implement and evaluate two real-world applications on \system{}: machine learning prediction serving and a Twitter clone ($\S$\ref{sec:eval-case}).

All experiments were run in the \texttt{us-east-1a} AWS availability zone.
Schedulers were run on AWS \texttt{c5.large} EC2 VMs (2 vCPUs, 4GB RAM), and function executors were run on \texttt{c5.2xlarge} EC2 VMs (8 vCPUs, 16GB RAM); 2 vCPUs comprise one physical core.
Function execution VMs used 3 cores for Python execution and 1 for the cache.
Clients were run on separate machines in the same AZ.
All Redis experiments were run using AWS Elasticache, using a cluster with two shards and three replicas per shard.

\subsection{Mechanisms in \system{}} \label{sec:eval-microbench}

In this section, we evaluate the primary individual mechanisms that Cloudburst enables---namely, low-latency function composition ($\S$\ref{sec:eval-compostition}), local cache data accesses ($\S$\ref{sec:eval-locality}), direct communication ($\S$\ref{sec:eval-comm}), and responsive autoscaling ($\S$\ref{sec:eval-scaling}).

\subsubsection{Function Composition} \label{sec:eval-compostition}

To begin, we compare \system{}'s function composition overheads with other serverless systems, as well as a non-serverless baseline.
We chose functions with minimal computation to isolate each system's overhead.
The pipeline was composed of two functions: \texttt{square(increment(x:int))}.
Figure~\ref{fig:e2e-latency} shows median and 99th percentile measured latencies across 1,000 requests run in serial from a single client.

First, we compare \system{} and Lambda using a ``stateless'' application, where we invoke one function---both bars are labelled stateless in Figure~\ref{fig:e2e-latency}.
\system{} stored results in Anna, as discussed in Section~\ref{sec:programming}. 
We ran \system{} with one function executor (3 worker threads).
We find that \system{} is about 5$\times$ faster than Lambda for this simple baseline.

For a composition of two functions---the simplest form of statefulness we support---we find that \system{}'s latency is roughly the same as with a single function and significantly faster than all other systems measured.
We first compared against \sand{}~\cite{akkus2018sand}, a new serverless platform that achieves low-latency function composition by using a hierarchical message bus.
We could not deploy \sand{} ourselves because the source code is unavailable, so we used the authors' hosted offering~\cite{sandserverless}.
As a result, we could not replicate the setup we used for the other experiments, where the client runs in the same datacenter as the service.
To compensate for this discrepancy, we accounted for the added client-server latency by measuring the latency for an empty HTTP request to the \sand{} service. 
We subtracted this number from the end-to-end latency for a request to our two-function pipeline running \sand{} to estimate the in-datacenter request time for the system.
In this setting, \sand{} is about an order of magnitude slower than \system{} both at median and at the 99th percentile.

To further validate \system{}, we compared against Dask, a ``serverful'' open-source distributed Python execution framework.
We deployed Dask on AWS using the same instances used for \system{} and found that performance was comparable to \system{}'s.
Given Dask's relative maturity, this gives us confidence that the overheads in Cloudburst are reasonable.

We compared against four AWS implementations, three of which used AWS Lambda.
Lambda (Direct) returns results directly to the user, while Lambda (S3) and Lambda (Dynamo) store results in the corresponding storage service.
All Lambda implementations pass arguments using the user-facing Lambda API.
The fastest implementation was Lambda (Direct) as it avoided high-latency storage, while DynamoDB added a 15ms latency penalty and S3 added 40ms.
We also compared against AWS Step Functions, which constructs a DAG of operations similar to \system{}'s and returns results directly to the user in a synchronous API call.
The Step Functions implementation was 10$\times$ slower than Lambda and 82$\times$ slower than \system{}.

\textit{\textbf{Takeaway}: \system{}'s function composition matches state-of-the-art Python runtime latency and outperforms commercial serverless infrastructure by 1-3 orders of magnitude.}

\subsubsection{Data Locality} \label{sec:eval-locality}

\begin{figure}[t]
  \centering
    \includegraphics[width=0.48\textwidth]{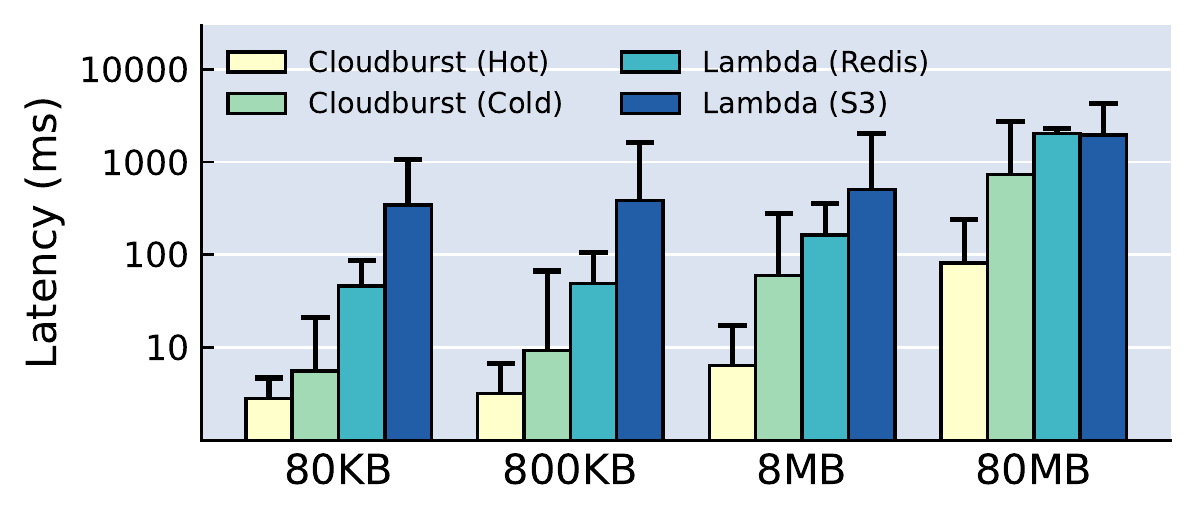}
  \caption{Median and 99th percentile latency to calculate the sum 10 arrays, comparing \system{} with caching and without and AWS Lambda over Redis and AWS S3. 
  We vary array lengths from 1,000 to 1,000,000 by multiples of 10 to demonstrate the effects of increasing data retrieval costs.}
  \vspace{-1em}
  \label{fig:locality}
\end{figure}

Next, we study the performance benefit of \system{}'s caching techniques.
We chose a representative task, with large input data but light computation: our function returns the sum of all elements across 10 input arrays.
We implemented two versions on AWS Lambda, which retrieved inputs from AWS ElastiCache (using Redis) and AWS S3 respectively.
ElastiCache is not an autoscaling system, but we include it in our evaluation because it offers best-case latencies for data retrieval for AWS Lambda.
We compare two implementations in \system{}.
One version, \system{} (Hot) passes the \emph{same} array in to every function execution, guaranteeing that every retrieval after the first is a cache hit.
This achieves optimal latency, as every request after the first avoids fetching data over the network.
The second, \system{} (Cold), creates a new set of inputs for each request; every retrieval is a cache miss, and this scenario measures worst-case latencies of fetching data from Anna.
All measurements are reported across 12 clients issuing 3,000 requests each.
We run \system{} with 7 function execution nodes.

The \system{} (Hot) bars in Figure~\ref{fig:locality} show that system's performance is consistent across the first two data sizes for cache hits, rises slightly for 8MB of data, and degrades significantly for the largest array size as computation costs begin to dominate.
\system{} performs best at 8MB, improving over \system{} (Cold)'s median latency by about 10$\times$, over Lambda on Redis' by 25$\times$, and over Lambda on S3's by 79$\times$.

While Lambda on S3 is the slowest configuration for smaller inputs, it is more competitive at 80MB.
Here, Lambda on Redis' latencies rise significantly.
\system{} (Cold)'s median latency is the second fastest, but its 99th percentile latency is comparable with S3's and Redis'.
This validates the common wisdom that S3 is efficient for high bandwidth tasks but imposes a high latency penalty for smaller data objects.
However, at this size, \system{} (Hot)'s median latency is still 9$\times$ faster than \system{} (Cold) and 24$\times$ faster than S3's.

\textit{\textbf{Takeaway}: While performance gains vary across configurations and data sizes, avoiding network roundtrips to storage services enables \system{} to improve performance by 1-2 orders of magnitude.}

\subsubsection{Low-Latency Communication} \label{sec:eval-comm}

\begin{figure}[t]
  \centering
    \includegraphics[width=\figwidth]{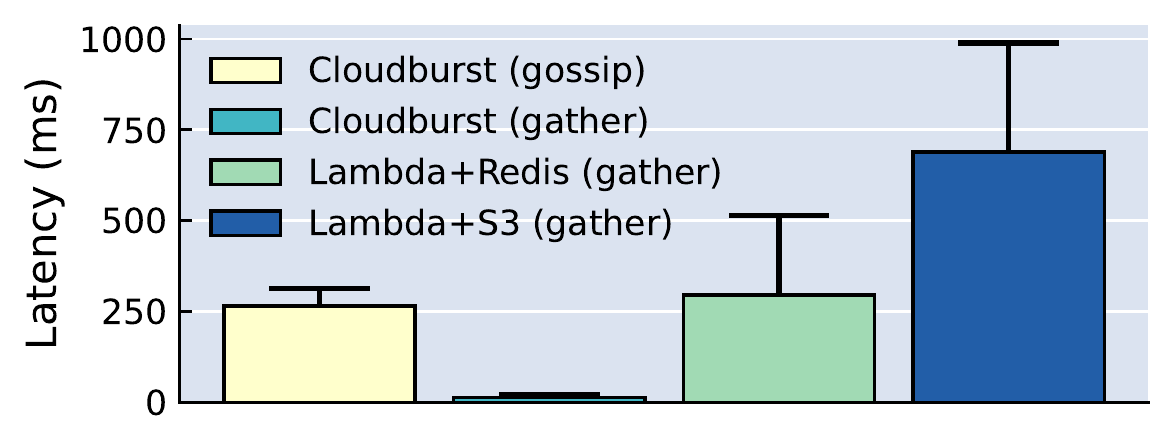}
  \caption{Median and 99th percentile latencies for distributed aggregation.
  The \system{} implementation uses a distributed, gossip-based aggregation technique, and the Lambda implementations share state via the respective key-value stores.
  \system{} outperforms communication through storage, even for a low-latency KVS.
  }
  \vspace{-1em}
  \label{fig:llcomm}
\end{figure}

Another key feature in \system{} is low-latency communication, which allows developers to leverage distributed systems protocols that are infeasibly slow in other serverless platforms~\cite{serverless-cidr19}.

As an illustration, we consider distributed aggregation, the simplest form of distributed statistics.
Our scenario is to periodically average a floating-point performance metric across the set of functions that are running at any given time. 
Kempe et al.~\cite{kempe2003gossip} developed a simple gossip-based protocol for approximate aggregation that uses random message passing among the current participants in the protocol. 
The algorithm is designed to provide correct answers even as the membership changes.
We implemented the algorithm in 60 lines of Python and ran it over \system{} with 4 executors (12 threads).
We compute 1,000 rounds of aggregation with 10 actors each in sequence and measure the time until the result converges to within 5\% error.

The gossip algorithm involves repeated small messages, making it highly inefficient on stateless platforms like AWS Lambda.
Since AWS Lambda disables direct messaging, the gossip algorithm would be extremely slow if implemented via reads/writes from slow storage.
Instead, we compare against a more natural approach for centralized storage: Each lambda function publishes its metrics to a KVS, and a predetermined leader gathers the published information and returns it to the client. 
We refer to this algorithm as the ``gather'' algorithm.
Note that this algorithm, unlike~\cite{kempe2003gossip}, requires the population to be fixed in advance, and is therefore not a good fit to an autoscaling setting. 
But it requires less communication, so we use it as a workaround to enable the systems that forbid direct communication to compete.
We implement the centralized gather protocol on Lambda over Redis for similar reasons as in $\S$~\ref{sec:eval-locality}---although serverful, Redis offers best-case performance for Lambda.
We also implement this algorithm over \system{} and Anna for reference.

Figure~\ref{fig:llcomm} shows our results.
\system{}'s gossip-based protocol is 3$\times$ faster than the gather protocol using Lambda and DynamoDB.
Although we expected gather on serverful Redis to outperform \system{}'s gossip algorithm, our measurements show that gossip on \system{} is actually about 10\% faster than the gather algorithm on Redis at median and 40\% faster at the 99th percentile.
Finally, gather on \system{} is 22$\times$ faster than gather on Redis and 53$\times$ faster than gather on DynamoDB.
There are two reasons for these discrepancies.
First, Lambda has very high function invocation costs (see $\S$\ref{sec:eval-compostition}).
Second, Redis is single-mastered and forces serialized writes, creating a queuing delay for writes.

\textit{\textbf{Takeaway}: \system{}'s low latency communication mechanisms enable developers to build fast distributed algorithms with fine-grained communication.
These algorithms can have notable performance benefits over workarounds involving even relatively fast shared storage.}

\subsubsection{Autoscaling} \label{sec:eval-scaling}

\begin{figure}[t]
  \centering
    \includegraphics[width=.42\textwidth]{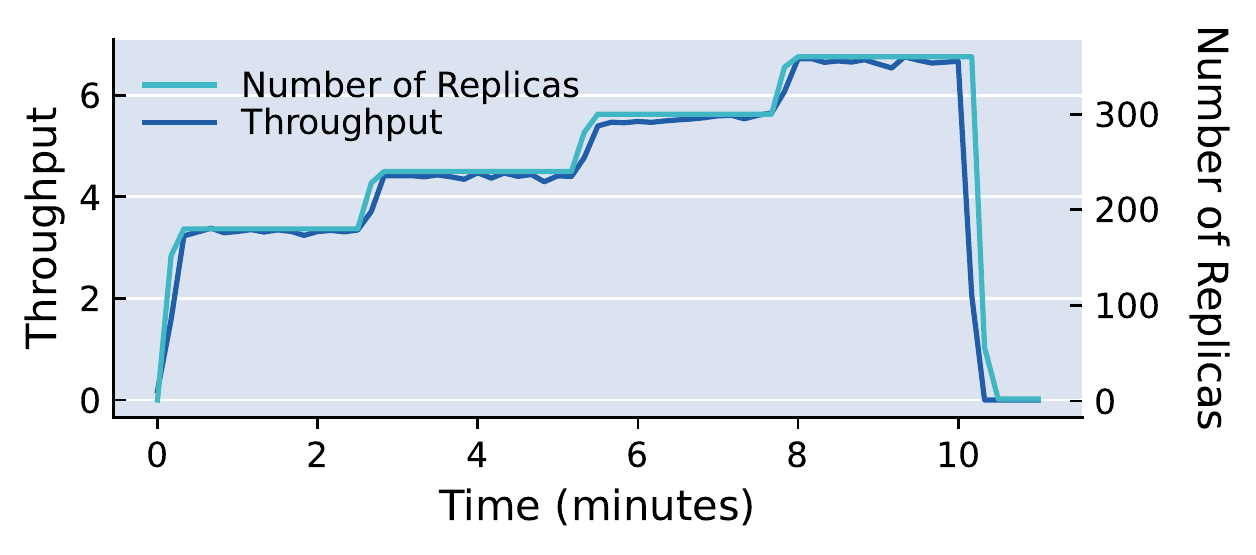}
  \caption{\system{}'s responsiveness to load changes. 
  We start with 180 executor threads, issue requests from 60 clients, and measure throughput. 
  \system{} quickly detects load spikes and allocate more resources. 
  Plateaus in the figure are the wait times for EC2 instance startup.} 
  \vspace{-1em}
  \label{fig:scaling}
\end{figure}

Finally, we validate \system{}'s ability to detect and respond to workload changes.
The goal of any serverless system is to smoothly scale program execution in response to changes in request rate.
As described in $\S$~\ref{sec:arch-fluent-mgmt}, \system{} uses a heuristic policy that accounts for incoming request rates, request execution times, and executor load.
We simulate a relatively computationally intensive workload with a function that sleeps for 50ms.
The function reads in two keys drawn from a Zipfian distribution with coefficient of 1.0 from 1 million 8-byte keys stored in Anna, and it writes to a third key drawn from the same distribution.

The system starts with 60 executors (180 threads) and one replica of the function deployed---the remaining threads are all idle.
Figure~\ref{fig:scaling} shows our results.
At time 0, 400 client threads simultaneously begin issuing requests.
The jagged curve measures system throughput (requests per second), and the dotted line tracks the number of threads allocated to the function.
Over the first 20 seconds, \system{} takes advantage of the idle resources in the system, and throughput reaches around 3,300 requests per second.
At this point, the management system detects that all nodes are saturated and adds 20 EC2 instances, which takes about 2.5 minutes; this is seen in the plateau that lasts until time 2.67.
As soon as resources become available, they are allocated to our task, and throughput rises to 4.4K requests a second.

This process repeats itself twice more, with the throughput rising to 5.6K and 6.7K requests per second with each increase in resources.
After 10 minutes, the clients stop issuing requests, and by time 10.33, the system has drained itself of all outstanding requests.
The management system detects the sudden drop in request rate and, within 20 seconds, reduces the number of threads allocated to the sleep function from 360 to 2.
Within 5 minutes, the number of EC2 instances drops from a max of 120 back to the original 60.
Our current implementation is bottlenecked by the latency of spinning up EC2 instances; we discuss that limitation and potential improvements in Section~\ref{sec:limitations}.

We also measured the per-key storage overhead of the index in Anna that maps each key to the caches it is stored in.
We observe small overheads even for the largest deployment (120 function execution nodes).
For keys in our working set, the median index overhead is 24 bytes and the 99th percentile overhead is 1.3KB, corresponding to keys being cached at 1.6\% and 93\% of the function nodes, respectively.
Even if all keys had the maximum overhead, the total index size would be around 1 GB for 1 million keys.

\textbf{Takeaway}: \system{}'s mechanisms for autoscaling enable policies that can quickly detect and react to workload changes.
We are mostly limited by the high cost of spinning up new EC2 instances. 
The policies and cost of spinning up instances can be improved in future without changing \system{}\'s architecture.

\subsection{Consistency Models} \label{sec:eval-consistency}

In this section, we evaluate the overheads of \system{}'s consistency models.
For comparison, we also implement and measure weaker consistency models to understand the costs involved in distributed session causal consistency. 
In particular, we evaluated \emph{single-key causality} and \emph{multi-key causality}, which are both weaker forms of causal consistency than the distributed session causality supported by \system{}.
\textit{Single-key causality} tracks causal order of updates to \emph{individual} keys (omitting the overhead of dependency sets).  
\textit{Multi-key causality} is an implementation of Bolt-On Causal Consistency~\cite{Bailis:2013:BCC:2463676.2465279}, avoiding the overhead of distributed session consistency.

We populate Anna with 1 million keys, each with a payload of 8 bytes, and we generate 250 random DAGs which are 2 to 5 functions long, with an average length of 3. 
We isolate the latency overheads of each consistency mode by avoiding expensive computation and use small data to highlight any metadata overheads.
Each function takes two string arguments, performs a simple string manipulation, and outputs another string.
Function arguments are either KVS references---which are drawn from the set of 1 million keys with a Zipfian coefficient of 1.0---or the result of a previous function execution in the DAG.
The sink function of the DAG writes its result to the KVS into a key chosen randomly from the set of keys read by the DAG.
We use 8 concurrent benchmark threads, each sequentially issuing 500 requests to \system{}.
We ran \system{} with 5 executor nodes (15 threads).

\subsubsection{Latency Comparison}

\begin{figure}[t]
  \centering
    \includegraphics[width=\figwidth]{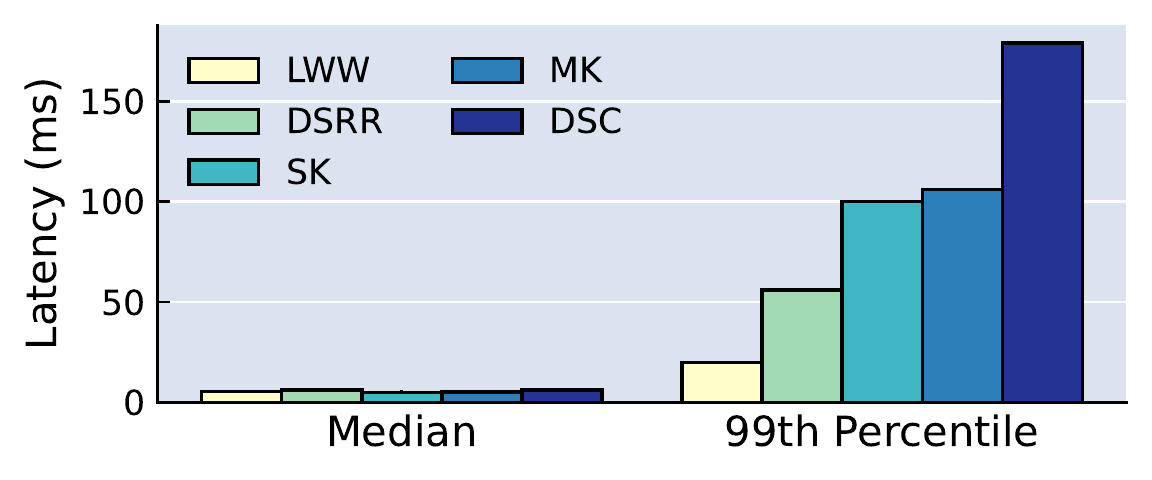}
  \caption{Median and 99th percentile latencies for \system{}'s consistency models, normalized by the depth of the DAG. 
  We measure last-writer wins (LWW), distributed session repeatable read (DSRR), single-key causality (SK), multi-key causality (MK), and distributed session causal consistency (DSC). 
  Median latency is uniform across modes, but stronger consistency levels have higher tail latencies due to increased data and metadata overheads.
  }
  \vspace{-0.5em}
  \label{fig:consistency-latency}
\end{figure}

Figure~\ref{fig:consistency-latency} shows the latency of each DAG execution under five consistency models normalized by the longest path in the DAG.
Median latency is nearly uniform across all modes, but performance differs significantly at the 99th percentile.

Last-writer wins has the lowest overhead, as it only stores the 8-byte timestamp associated with each key and requires no remote version fetches.
The 99th percentile latency of distributed session repeatable read is 1.8$\times$ higher than last-writer wins'. 
This is because repeated reference to a key across functions requires an \emph{exact} version match; even if the key is cached locally, a version mismatch will force a remote fetch.

Single-key causality does not involve metadata passing or data retrieval, but each key maintains a vector clock that tracks the causal ordering of updates performed across clients.
Since the size of the vector clock grows linearly with the number of clients that modified the key, hot keys tend to have larger vector clocks, leading to higher retrieval latency at the tail.
Multi-key causality forces each key to track its dependencies in addition to maintaining the vector clock, adding to its worst-case latency.
We observe that the median per-key causal metadata (vector clock and dependency set) storage overhead is 624 bytes and the 99th percentile overhead is 7.1KB.
Techniques such as vector clock compression~\cite{akbar2017icant} and dependency garbage collection~\cite{COPS} can be used to reduce the metadata overhead, which we plan to explore in future work.

Distributed session causal consistency incurs the cost of passing causal metadata along the DAG as well as retrieving version snapshots to satisfy causality.
In the worst case, a 5-function DAG performs 4 extra network round-trips for version snapshots.
This leads to a 1.7$\times$ slowdown in 99th percentile latency over single- and multi-key causality and a 9$\times$ slowdown over last-writer wins.

\textit{\textbf{Takeaway}: Although \system{}'s non-trivial consistency models increase tail latencies, median latencies are over an order of magnitude faster than DynamoDB and S3 for similar tasks, while providing stronger consistency.}

\subsubsection{Inconsistencies}

\renewcommand{\arraystretch}{1.4}
\begin{table}\scriptsize
\caption{The number of inconsistencies observed across 4,000 DAG executions under \system{}'s consistency levels relative to what is observed under LWW.
The causal levels are increasingly strict, so the numbers accrue incrementally left to right. 
DSRR anomalies are independent.
}
\begin{center}
\footnotesize
\begin{tabular}{|c|c|c|c|c|}
\cline{1-5}
\multicolumn{5}{|c|}{Inconsistencies Observed} \\
\cline{1-5}
LWW & \multicolumn{3}{|c|} {Causal} & DSRR \\
\cline{2-4}
 & SK & MK & DSC &  \\
\hline
0 & 904 & 939 & 1043 & 46 \\
\hline
\end{tabular}
\end{center}
\vspace{-1em}
\label{fig:consistency-inconsistency}
\end{table}


Stronger consistency models introduce overheads but also prevent anomalies that would arise in weaker models.
Table~\ref{fig:consistency-inconsistency} shows the number of anomalies observed over the course of 4000 DAG executions run in LWW mode, tracking anomalies for other levels.

The causal consistency levels have increasingly strict criteria; anomaly counts accrue with the level.
We observe 904 single-key (SK) causal inconsistencies when the system operates in LWW mode.
With single-key causality, two concurrent updates to the same must both be preserved and returned to the client.
LWW simply picks the largest timestamp and drops the other update, leading to a majority of observed anomalies.
Multi-key (MK) causality flagged 35 additional inconsistencies corresponding to single-cache read sets that were not causal cuts.
Distributed session causal consistency (DSC) flagged 104 more inconsistencies where the causal cut property was violated across caches.
Repeatable Read (DSRR) flagged 46 anomalies.

%

\textit{\textbf{Takeaway}:
A large number of anomalies arose naturally in our experiments and the \system{} consistency model was able to detect and prevent these anomalies.
}

\subsection{Case Studies} \label{sec:eval-case}

In this section, we discuss the implementation of two real-world applications on top of \system{}.
We first consider low-latency prediction serving for machine learning models and compare \system{} to a purpose-built cloud offering, AWS Sagemaker.
We then implement a Twitter clone called Retwis, which takes advantage of our consistency mechanisms, and we report both the effort involved in porting the application to \system{} as well as some initial evaluation metrics.

\subsubsection{Prediction Serving} \label{sec:eval-serving}

\begin{figure}[t]
  \centering
    \includegraphics[width=\figwidth]{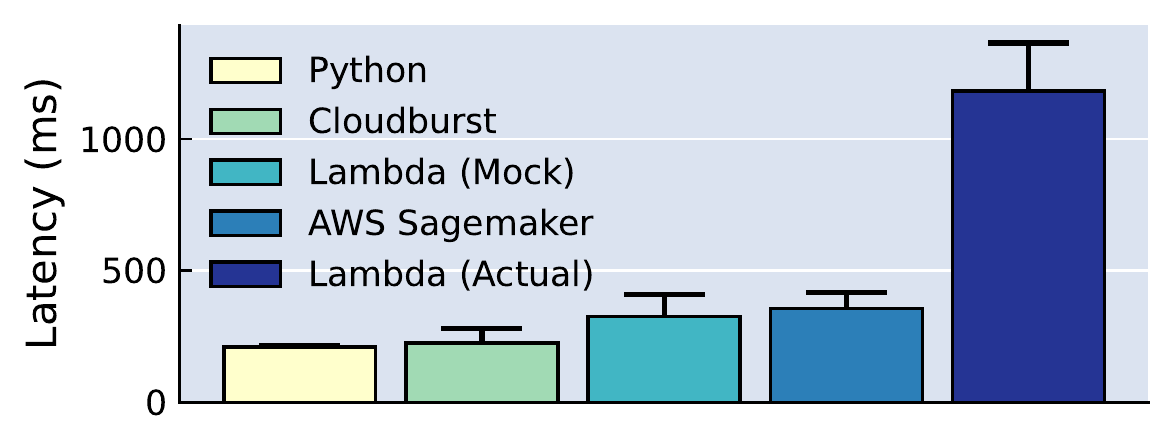}
  \caption{\system{} compared to native Python, AWS Sagemaker, and AWS Lambda for serving a prediction pipeline.}
  \vspace{-0.5em}
  \label{fig:pred-serving}
\end{figure}

\begin{figure}[t]
  \centering
    \includegraphics[width=\figwidth]{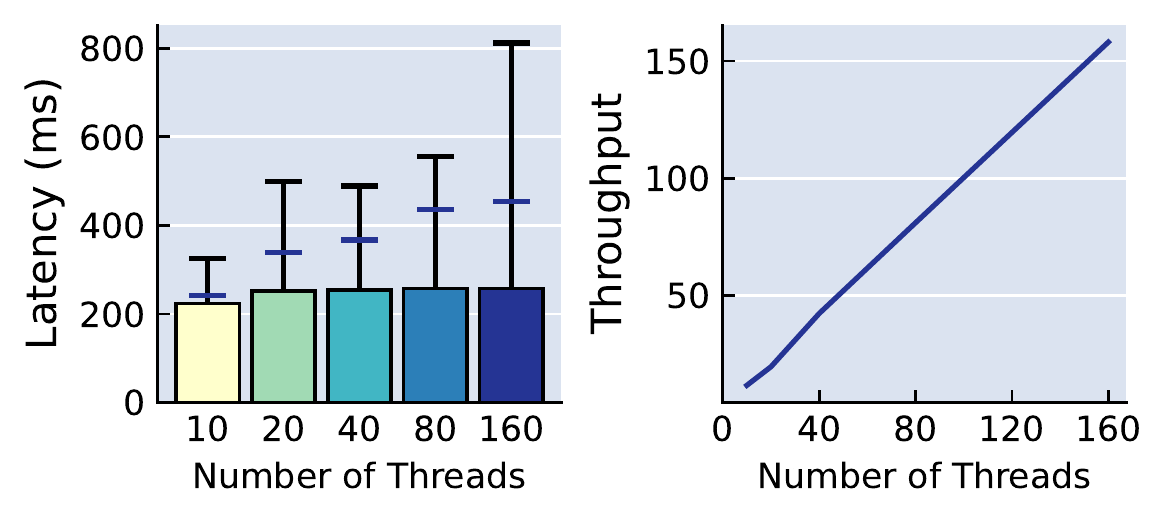}
  \caption{A measure of the \system{}'s ability to scale a simple prediction serving pipeline.
  The blue whiskers represent 95th percentile latencies, and the black represent the 99th percentile.}
  \vspace{-1em}
  \label{fig:pred-serving-scale}
\end{figure}

ML model prediction is a computationally intensive task that can benefit from elastic scaling and efficient sparse access to large amounts of state.
For example, the prediction serving infrastructure at Facebook~\cite{facebook} needs to access per-user state with each query and respond in real time, with strict latency constraints.
Furthermore, many prediction pipelines combine multiple stages of computation---e.g., clean the input, join it with reference data, execute one or more models, and combine the results~\cite{clipper-nsdi17, pretzel}.

We implemented a basic prediction serving pipeline on \system{} and compare against a fully-managed, purpose-built prediction serving framework (AWS Sagemaker) as well as AWS Lambda.
We also compare against a single Python process to measure serialization and communication overheads.
Lambda does not support GPUs, so all experiments are run on CPUs.

We use the MobileNet~\cite{mobilenet} image classification model implemented in Tensorflow~\cite{abadi2016tensorflow} and construct a three-stage pipeline: resize an input image, execute the model, and combine features to render a prediction.
Qualitatively, porting this pipeline to \system{} was easier than porting it to other systems.
The native Python implementation was 23 lines of code (LOC).
\system{} required adding 4 LOC to retrieve the model from Anna.
AWS SageMaker required adding serialization logic (10 LOC) and a Python web-server to invoke each function (30 LOC).
Finally, AWS Lambda required significant changes: managing serialization (10 LOC) and manually compressing Python dependencies to fit into Lambda's 512MB container limit\footnote{AWS Lambda limits disk space to 512MB. Tensorflow exceeds this limit, so we removed unnecessary components. 
We do not report LOC changed as it would artificially inflate the estimate.}.
The pipeline does not involve concurrent modification to shared state, so we use the default last writer wins for this workload.
We run \system{} with 3 workers, and all experiments used a single client.

Figure~\ref{fig:pred-serving} reports median and 99th percentile latencies.
\system{} is only about 15ms slower than the Python baseline at the median (210ms vs. 225ms).
AWS Sagemaker, ostensibly a purpose-built system, is 1.7$\times$ slower than the native Python implementation and 1.6$\times$ slower than \system{}.
We also measure two AWS Lambda implementations.
One, AWS Lambda (Actual), computes a full result for the pipeline and takes over 1.1 seconds.
To better understand Lambda's performance, we isolated compute costs by removing all data movement. 
This result (AWS Lambda (Mock)) is much faster, suggesting that the latency penalty is incurred by the Lambda runtime passing results between functions.
Nonetheless, AWS Lambda (Mock)'s median is still 44\% slower than \system{}'s median latency and only 9\% faster than AWS Sagemaker.

Figure~\ref{fig:pred-serving-scale} measures throughput and latency for \system{} as we increase the number of worker threads from 10 to 160 by factors of two. 
The number of clients for each setting is set to $ \lfloor \frac{workers}{3} \rfloor$ because there are three functions executed per client.
We see that throughput scales linearly with the number of workers. 
We see a climb in median and 99th percentile latency from 10 to 20 workers due to increased potential conflicts in the scheduling heuristics.
From this point on, we do not a significant change in either median or tail latency until 160 executors.
For the largest deployment, only one or two executors need to be slow to significantly raise the 99th percentile latency---to validate this, we also report the 95th percentile latency in Figure~\ref{fig:pred-serving-scale}, and we see that there is a minimal increase between 80 and 160 executors.

\emph{\textbf{Takeaway:}} \textit{An ML algorithm deployed on \system{} delivers both smooth scaling and low, predictable latency comparable to native Python, out-performing a purpose-built commercial service.} 

\subsubsection{Retwis} \label{sec:eval-retwis}

\begin{figure}[t]
  \centering
    \includegraphics[width=\figwidth]{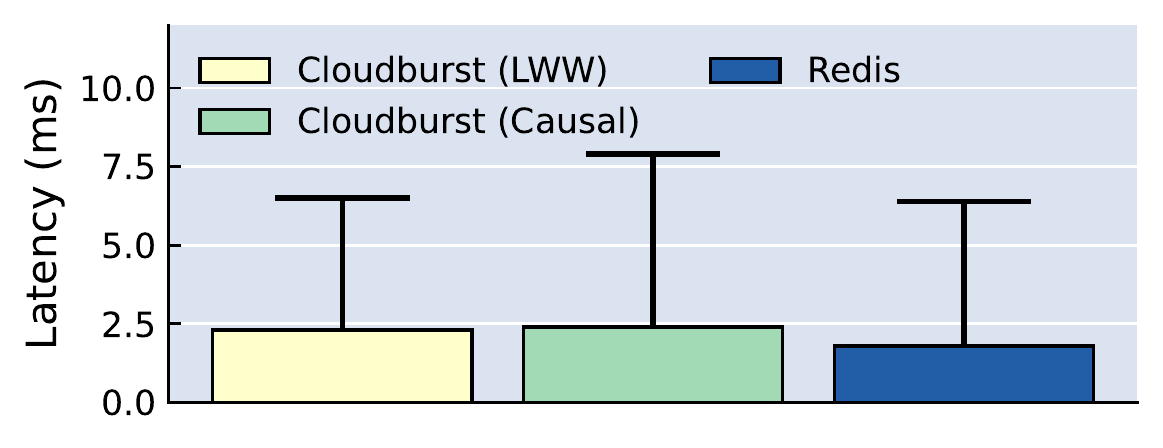}
  \caption{
    Median and 99th percentile latencies for \system{} in LWW and causal modes and Retwis over Redis.
  }
  \vspace{-0.5em}
  \label{fig:retwis}
\end{figure}

\begin{figure}[t]
  \centering
    \includegraphics[width=\figwidth]{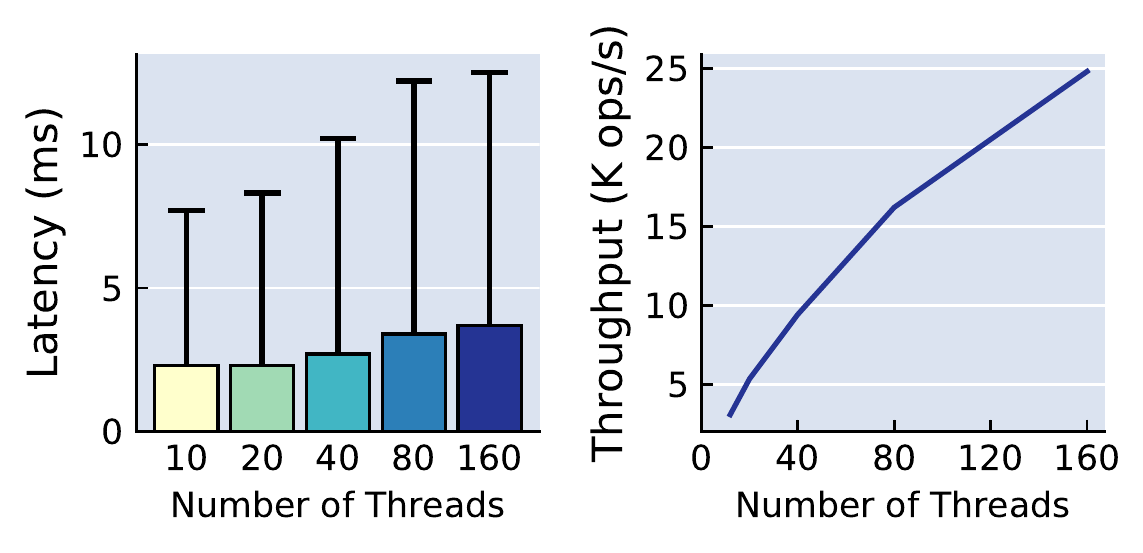}
  \caption{\system{}'s ability to scale the Retwis workload up to 160 worker threads.}
  \vspace{-1em}
  \label{fig:twitter-scale}
\end{figure}

Web serving workloads are closely aligned with \system{}'s features.
For example, Twitter provisions server capacity of up to 10x the typical daily peak in order to accommodate unusual events such as elections, sporting events, or natural disasters~\cite{twitterinfra}.
Furthermore, causal consistency is a good model for many consumer internet workloads because it matches end-user expectations for information propagation: e.g., Google has adopted it as part of a universal model for privacy and access control~\cite{pang2019zanzibar}.

To this end, we considered an example web serving workload.
Retwis~\cite{retwis} is an open source Twitter clone built on Redis and is often used to evaluate distributed systems~\cite{sovran2011transactional,holt2015claret,zhang2016diamond,crooks2016tardis,yan2018carousel}.
Conversational ``threads'' like those on Twitter naturally exercise causal consistency: It is confusing to read the response to a post (e.g., ``lambda!'') before you have read the post it refers to (``what comes after kappa?'').
We adapted a Python Retwis implementation called \texttt{retwis-py}~\cite{retwis-py} to run on \system{} and compared its performance to a vanilla ``serverful'' deployment on Redis.
We ported Retwis to our system as a set of six \system{} functions.
The port was simple: We changed 44 lines, most of which were removing references to a global Redis variable.

We created a graph of 1000 users, each following 50 other users (zipf=1.5, a realistic skew for online social networks~\cite{social-network-zipf}) and prepopulated 5000 tweets, half of which were replies to other tweets.
We compare \system{} in LWW mode, \system{} in causal consistency mode, and Retwis over Redis; all configurations used 10 executor threads (webservers for Retwis), 1 KVS node, and 10 clients.
Each client issues 5000 requests---10\% \texttt{PostTweet} (write) requests and 90\% \texttt{GetTimeline} (read) requests.

Figure~\ref{fig:retwis} shows our results.
Median and 99th percentile latencies for LWW mode are 27\% and 2\% higher than Redis', respectively.
This is largely due to different code paths; for Retwis over Redis, clients communicate directly with web servers, which interact with Redis.
Each \system{} request interacts with a scheduler, a function executor, a cache, and Anna.
\system{}'s causal mode adds a modest overhead over LWW mode: 4\% higher at the median and 20\% higher at the tail.
However, causality prevents anomalies on over 60\% of requests---when a timeline returns a reply without the original tweet---compared to LWW mode.

Figure~\ref{fig:twitter-scale} measures throughput and latency for \system{}'s causal mode as we increase the number of function executor threads from 10 to 160 by factors of two. 
For each setting, the number of clients is equal to the number of executors.
From 10 threads to 160 threads, median and the 99th percentile latencies increase by about 60\%.
This is because increased concurrency means a higher volume of new tweets.
With more new posts, each \texttt{GetTimeline} request forces the cache to query the KVS for new data with higher probability in order to ensure causality---for 160 threads, 95\% of requests incurred cache misses.
Nonetheless, these latencies are well within the bounds for interactive web applications~\cite{he2012zeta}.
Throughput grows nearly linearly as we increase the executor thread count.
However, due to the increased latencies, throughput is about 30\% below ideal at the largest scale.

\emph{\textbf{Takeaway:}} \textit{It was straightforward to adapt a standard social network application to run on \system{}.
Our implementation adds a modest overhead to the serverful Redis baseline and scales smoothly as the workload increases.}

%% file: tex/related.tex
\vspace{-0.5em}
\section{Related Work} \label{sec:related}

\smallitem{Architecture}. Client-side caching and consistency has a long history in the database literature~\cite{franklin1992client} that bears similarity to our LDPC principal.
In more recent examples, \cite{brantner2008building} describes client-side techniques for a database built on S3, and \cite{zamanian2016end} uses caching in a distributed transactional system based on RDMA over dedicated memory nodes.
This line of work primarily focuses on strong transactional consistency for static or slowly-changing configurations.
\system{} explicitly chooses to pursue coordination-free techniques for serverless computing, for the reasons described in Section~\ref{sec:motiv-challenges}. 
The serverless setting also drives our need to handle consistency when a single transaction moves across multiple caches, which is not seen in prior work.

\smallitem{Serverless Execution Frameworks}.
In addition to commercial offerings, there are many open-source serverless platforms~\cite{openlambda, openwhisk, openfaas, kubeless} which provide standard stateless FaaS guarantees.
Among research platforms~\cite{hendrickson2016serverless,akkus2018sand,mcgrath2017serverless}, \texttt{SAND}~\cite{akkus2018sand} is most similar to \system{}, reducing overheads for low-latency function compositions.
\system{} achieves better latencies ($\S$\ref{sec:eval-compostition}) and adds shared state and communication abstractions. 

Recent work has explored faster, more efficient serverless platforms.
SOCK~\cite{sock-atc18} introduces a generalized-Zygote provisioning mechanism to cache and clone function initialization; its library loading technique could be integrated with \system{}.
Also complementary are low-overhead sandboxing mechanisms released by cloud providers---e.g., gVisor~\cite{gvisor} and Firecracker~\cite{firecracker}.

Other recent work has demonstrated the ability to build data-centric services on top of commodity serverless infrastructure.
Starling~\cite{perron2019starling} implements a scalable database query engine on AWS Lambda.
Similarly, the PyWren project~\cite{pywren} and follow-on work~\cite{numpywren} implemented highly parallel map tasks on Lambda, with a focus on linear algebra applications.
The ExCamera project~\cite{excamera} enabled highly efficient video encoding on FaaS.
These systems explored what is possible to build on stateless serverless infrastructure
By contrast, our work explores the design of a new architecture, which raises new challenges for systems issues in distributed
consistency.

Perhaps the closest work to ours is the Archipelago system~\cite{archipelago}, which also explores the design of a new serverless infrastructure.
They also adopt the model of DAGs of functions, but their focus is complementary to ours.
They are interested in scheduling techniques to achieve latency deadlines on a per-request basis.

\smallitem{Serverless IO Infrastructure}. Pocket~\cite{pocket} is a serverless storage system that improves the efficiency of analytics (large read oriented workloads).
Locus~\cite{pu2019shuffling} provides a high-bandwidth shuffle functionality for serverless analytics.
Neither of these systems offers a new serverless programming framework, nor do they consider issues of caching for latency or consistent updates.

Finally, Shredder~\cite{zhang2019narrowing} enables users to push functions into storage.
This raises fundamental security concerns like multi-tenant isolation, which are the focus of the research.
Shredder is currently limited to a single node and does not address autoscaling or other characteristic features of serverless computing.

\smallitem{Language-Level Consistency} 
Programming languages also offer solutions to distributed consistency.
One option from functional languages is to \emph{prevent} inconsistencies by making state \emph{immutable}~\cite{coblenz2016exploring}. 
A second is to constrain updates to be \emph{deterministically mergeable}, by requiring users to write associative, commutative, and idempotent code (ACID 2.0~\cite{campbell}), use special-purpose types like CRDTs~\cite{shapiro2011conflict} or DSLs for distributed computing like Bloom~\cite{Conway:2012:LLD:2391229.2391230}.
As a platform, \system{} does not prescribe a language or type system, though these approaches could be layered on top of \system{}.

\smallitem{Causal Consistency}.
Several existing storage systems provide causal consistency~\cite{COPS, akbar2017icant, akkoorath2016cure, lloyd2013stronger, du2014gentlerain, Du:2013:OSC:2523616.2523628, Almeida:2013:CCC:2465351.2465361, Zawirski:2015:WFR:2814576.2814733}. 
However, these are fixed-deployment systems that do not meet the autoscaling requirements of a serverless setting.
In ~\cite{akbar2017icant, akkoorath2016cure, Almeida:2013:CCC:2465351.2465361, Du:2013:OSC:2523616.2523628, du2014gentlerain}, each data partition relies on a linear clock to version data and uses a fixed-size vector clock to track causal dependencies across keys.
The size of these vector clocks is tightly coupled with the system deployment---specifically, the shard and replica counts.
Correctly adjusting this metadata requires an expensive coordination protocol, which we rejected in \system{}'s design ($\S$~\ref{sec:motiv-challenges}).
\cite{COPS} and \cite{lloyd2013stronger} reveal a new version only when all of its dependencies have been retrieved.
\cite{Zawirski:2015:WFR:2814576.2814733} constructs a causally consistent snapshot across an entire data center.
All of these systems are susceptible to ``slowdown cascades''~\cite{akbar2017icant}, where a single straggler node limits write visibility and increases the overhead of write buffering.

In contrast, \system{} implements causal consistency in the cache layer as in Bolt-On Causal Consistency~\cite{Bailis:2013:BCC:2463676.2465279}.
Each cache creates its own causally consistent snapshot without coordinating with other caches, eliminating the possibility of a slowdown cascade.
The cache layer also tracks dependencies in individual keys' metadata rather than tracking the vector clocks of fixed, coarse-grained shards.
This comes at the cost of increased dependency metadata overhead. 
Various techniques including periodic dependency garbage collection~\cite{COPS}, compression~\cite{akbar2017icant}, and reducing dependencies via explicit causality specification~\cite{Bailis:2013:BCC:2463676.2465279} can mitigate this issue, though we do not measure them here.

%% file: tex/conclusion.tex
\section{Conclusion and Future Work} \label{sec:conclusion}

In this paper we demonstrate the feasibility of general-purpose 
stateful serverless computing.  
We enable autoscaling via logical disaggregation of storage and compute and achieve performant state management via physical colocation of caches with compute services.
\system{} demonstrates that disaggregation and colocation are not inherently in conflict. 
In fact, the LDPC design pattern is key to our solution for stateful serverless computing. 

The remaining challenge is to provide performant correctness. 
\system{} 
embraces coordination-free consistency
as the appropriate class of guarantees for an autoscaling system. We
confront challenges at both
the storage and caching layer. We use lattice capsules to 
allow opaque program state to be merged asynchronously into replicated coordination-free persistent storage.
We develop distributed session consistency protocols to ensure that computations spanning multiple caches 
provide uniform correctness guarantees. Together, these techniques provide a strong contract to users
for reasoning about state---far stronger than the guarantees offered by cloud storage that backs commercial FaaS systems.
Even with these guarantees, we demonstrate performance that rivals and often beats baselines from inelastic server-centric approaches. 

\input{tex/limits.tex}


%% file: tex/limits.tex
\label{sec:limitations}
The feasibility of stateful serverless computing suggests a variety of potential future work. 

\smallitem{Isolation and Fault Tolerance}
As noted in $\S$\ref{sec:consistency-guarantees}, storage consistency guarantees say nothing about concurrent effects \emph{between} DAGs, a concern akin to transactional isolation (the ``I'' in ACID).
On a related note, the standard fault tolerance model for FaaS is to restart on failure, ignoring potential problems with non-idempotent functions. 
Something akin to transactional atomicity (the ``A'' in ACID) seems desirable here.
It is well-known that serializable transactions require coordination~\cite{Bailis:2013:HAT:2732232.2732237}, but it is interesting to consider whether sufficient atomicity and isolation are achievable without strong consistency schemes.


\smallitem{Auto-Scaling Mechanism and Policy}.
In this work we present and evaluate a simple auto-scaling heuristic.
However, there are opportunities to reduce boot time by warm pooling~\cite{wagnercompute} and more proactively scale computation~\cite{autoscale, ds2, taft-thesis} as a function of variability in the workloads. 
We believe that cache-based co-placement of computation and data presents promising opportunities for research in elastic auto-scaling of compute and storage. 

\smallitem{Streaming Services}
\system{}'s internal monitoring service requires components to publish metadata to well-known KVS keys.
In essence, the KVS serves a rolling snapshot of an update stream. 
There is a rich literature on distributed streaming that could offer more, e.g. as surveyed by~\cite{garofalakis2016data}. 
The autoscaling environment of FaaS introduces new challenges, but this area seems ripe for both system internals and user-level streaming abstractions.

\smallitem{Security and Privacy}.
As mentioned briefly in $\S$\ref{sec:arch}, \system{}'s current design provides container-level isolation, which is susceptible to well-known attacks~\cite{meltdown, light-vm}.
This is unacceptable in multi-tenant cloud environments, where sensitive user data may coincide with other user programs.
It is interesting to explore how \system{}'s design would address these concerns.

%% file: main.bbl
\begin{thebibliography}{10}

\bibitem{abadi2016tensorflow}
M.~Abadi, P.~Barham, J.~Chen, Z.~Chen, A.~Davis, J.~Dean, M.~Devin,
  S.~Ghemawat, G.~Irving, M.~Isard, et~al.
\newblock Tensorflow: A system for large-scale machine learning.
\newblock In {\em 12th $\{$USENIX$\}$ Symposium on Operating Systems Design and
  Implementation ($\{$OSDI$\}$ 16)}, pages 265--283, 2016.

\bibitem{airflow}
Apache airflow.
\newblock \url{https://airflow.apache.org}.

\bibitem{akkoorath2016cure}
D.~D. Akkoorath, A.~Z. Tomsic, M.~Bravo, Z.~Li, T.~Crain, A.~Bieniusa,
  N.~Pregui{\c{c}}a, and M.~Shapiro.
\newblock Cure: Strong semantics meets high availability and low latency.
\newblock In {\em 2016 IEEE 36th International Conference on Distributed
  Computing Systems (ICDCS)}, pages 405--414. IEEE, 2016.

\bibitem{akkus2018sand}
I.~E. Akkus, R.~Chen, I.~Rimac, M.~Stein, K.~Satzke, A.~Beck, P.~Aditya, and
  V.~Hilt.
\newblock {SAND}: Towards high-performance serverless computing.
\newblock In {\em 2018 USENIX Annual Technical Conference (USENIX ATC 18)},
  pages 923--935, 2018.

\bibitem{Almeida:2013:CCC:2465351.2465361}
S.~Almeida, J.~a. Leit\~{a}o, and L.~Rodrigues.
\newblock Chainreaction: A causal+ consistent datastore based on chain
  replication.
\newblock In {\em Proceedings of the 8th ACM European Conference on Computer
  Systems}, EuroSys '13, pages 85--98, New York, NY, USA, 2013. ACM.

\bibitem{awerbuch1987optimal}
B.~Awerbuch.
\newblock Optimal distributed algorithms for minimum weight spanning tree,
  counting, leader election, and related problems.
\newblock In {\em Proceedings of the nineteenth annual ACM symposium on Theory
  of computing}, pages 230--240. ACM, 1987.

\bibitem{awscasestudies}
Aws {L}ambda - case studies.
\newblock \url{https://aws.amazon.com/lambda/resources/customer-case-studies/}.

\bibitem{Bailis:2013:HAT:2732232.2732237}
P.~Bailis, A.~Davidson, A.~Fekete, A.~Ghodsi, J.~M. Hellerstein, and I.~Stoica.
\newblock Highly available transactions: Virtues and limitations.
\newblock {\em PVLDB}, 7(3):181--192, 2013.

\bibitem{Bailis:2013:BCC:2463676.2465279}
P.~Bailis, A.~Ghodsi, J.~M. Hellerstein, and I.~Stoica.
\newblock Bolt-on causal consistency.
\newblock In {\em Proceedings of the 2013 ACM SIGMOD International Conference
  on Management of Data}, SIGMOD '13, pages 761--772, New York, NY, USA, 2013.
  ACM.

\bibitem{baldini2017serverless}
I.~Baldini, P.~Castro, K.~Chang, P.~Cheng, S.~Fink, V.~Ishakian, N.~Mitchell,
  V.~Muthusamy, R.~Rabbah, A.~Slominski, et~al.
\newblock Serverless computing: Current trends and open problems.
\newblock In {\em Research Advances in Cloud Computing}, pages 1--20. Springer,
  2017.

\bibitem{berenson1995critique}
H.~Berenson, P.~Bernstein, J.~Gray, J.~Melton, E.~O'Neil, and P.~O'Neil.
\newblock A critique of {ANSI} {SQL} isolation levels.
\newblock In {\em ACM SIGMOD Record}, volume~24, pages 1--10. ACM, 1995.

\bibitem{birman1987exploiting}
K.~Birman and T.~Joseph.
\newblock Exploiting virtual synchrony in distributed systems.
\newblock {\em SIGOPS Oper. Syst. Rev.}, 21(5):123--138, Nov. 1987.

\bibitem{brantner2008building}
M.~Brantner, D.~Florescu, D.~Graf, D.~Kossmann, and T.~Kraska.
\newblock Building a database on s3.
\newblock In {\em Proceedings of the 2008 ACM SIGMOD international conference
  on Management of data}, pages 251--264, 2008.

\bibitem{brewercap}
E.~{Brewer}.
\newblock Cap twelve years later: How the ``rules'' have changed.
\newblock {\em Computer}, 45(2):23--29, Feb 2012.

\bibitem{chandra2007paxos}
T.~D. Chandra, R.~Griesemer, and J.~Redstone.
\newblock Paxos made live: an engineering perspective.
\newblock In {\em Proceedings of the twenty-sixth annual ACM symposium on
  Principles of distributed computing}, pages 398--407. ACM, 2007.

\bibitem{coblenz2016exploring}
M.~Coblenz, J.~Sunshine, J.~Aldrich, B.~Myers, S.~Weber, and F.~Shull.
\newblock Exploring language support for immutability.
\newblock In {\em Proceedings of the 38th International Conference on Software
  Engineering}, pages 736--747. ACM, 2016.

\bibitem{Conway:2012:LLD:2391229.2391230}
N.~Conway, W.~R. Marczak, P.~Alvaro, J.~M. Hellerstein, and D.~Maier.
\newblock Logic and lattices for distributed programming.
\newblock In {\em Proceedings of the Third ACM Symposium on Cloud Computing},
  SoCC '12, pages 1:1--1:14, New York, NY, USA, 2012. ACM.

\bibitem{clipper-nsdi17}
D.~Crankshaw, X.~Wang, G.~Zhou, M.~J. Franklin, J.~E. Gonzalez, and I.~Stoica.
\newblock Clipper: A low-latency online prediction serving system.
\newblock In {\em 14th {USENIX} Symposium on Networked Systems Design and
  Implementation ({NSDI} 17)}, pages 613--627, Boston, MA, 2017. {USENIX}
  Association.

\bibitem{crooks2016tardis}
N.~Crooks, Y.~Pu, N.~Estrada, T.~Gupta, L.~Alvisi, and A.~Clement.
\newblock Tardis: A branch-and-merge approach to weak consistency.
\newblock In {\em Proceedings of the 2016 International Conference on
  Management of Data}, pages 1615--1628. ACM, 2016.

\bibitem{das2002swim}
A.~Das, I.~Gupta, and A.~Motivala.
\newblock Swim: Scalable weakly-consistent infection-style process group
  membership protocol.
\newblock In {\em Proceedings International Conference on Dependable Systems
  and Networks}, pages 303--312. IEEE, 2002.

\bibitem{docker}
Enterprise application container platform | docker.
\newblock \url{https://www.docker.com}.

\bibitem{Du:2013:OSC:2523616.2523628}
J.~Du, S.~Elnikety, A.~Roy, and W.~Zwaenepoel.
\newblock Orbe: Scalable causal consistency using dependency matrices and
  physical clocks.
\newblock In {\em Proceedings of the 4th Annual Symposium on Cloud Computing},
  SOCC '13, pages 11:1--11:14, New York, NY, USA, 2013. ACM.

\bibitem{du2014gentlerain}
J.~Du, C.~Iorgulescu, A.~Roy, and W.~Zwaenepoel.
\newblock Gentle{R}ain: Cheap and scalable causal consistency with physical
  clocks.
\newblock In {\em Proceedings of the ACM Symposium on Cloud Computing}, pages
  1--13. ACM, 2014.

\bibitem{firecracker}
Announcing the firecracker open source technology: Secure and fast micro{VM}
  for serverless computing.
\newblock
  \url{https://aws.amazon.com/blogs/opensource/firecracker-open-source-secure-fast-microvm}
  \url{-serverless/}.

\bibitem{fouladi2019laptop}
S.~Fouladi, F.~Romero, D.~Iter, Q.~Li, S.~Chatterjee, C.~Kozyrakis, M.~Zaharia,
  and K.~Winstein.
\newblock From laptop to {L}ambda: Outsourcing everyday jobs to thousands of
  transient functional containers.
\newblock In {\em 2019 USENIX Annual Technical Conference (USENIX ATC 19)},
  pages 475--488, 2019.

\bibitem{excamera}
S.~Fouladi, R.~S. Wahby, B.~Shacklett, K.~V. Balasubramaniam, W.~Zeng,
  R.~Bhalerao, A.~Sivaraman, G.~Porter, and K.~Winstein.
\newblock Encoding, fast and slow: Low-latency video processing using thousands
  of tiny threads.
\newblock In {\em 14th {USENIX} Symposium on Networked Systems Design and
  Implementation ({NSDI} 17)}, pages 363--376, Boston, MA, 2017. {USENIX}
  Association.

\bibitem{franklin1992client}
M.~J. Franklin and M.~J. Carey.
\newblock Client-server caching revisited.
\newblock Technical report, University of Wisconsin-Madison Department of
  Computer Sciences, 1992.

\bibitem{gan2019open}
Y.~Gan, Y.~Zhang, D.~Cheng, A.~Shetty, P.~Rathi, N.~Katarki, A.~Bruno, J.~Hu,
  B.~Ritchken, B.~Jackson, et~al.
\newblock An open-source benchmark suite for microservices and their
  hardware-software implications for cloud \& edge systems.
\newblock In {\em Proceedings of the Twenty-Fourth International Conference on
  Architectural Support for Programming Languages and Operating Systems}, pages
  3--18. ACM, 2019.

\bibitem{autoscale}
A.~Gandhi, M.~Harchol-Balter, R.~Raghunathan, and M.~A. Kozuch.
\newblock Autoscale: Dynamic, robust capacity management for multi-tier data
  centers.
\newblock {\em ACM Transactions on Computer Systems (TOCS)}, 30(4):14, 2012.

\bibitem{garofalakis2016data}
M.~Garofalakis, J.~Gehrke, and R.~Rastogi.
\newblock {\em Data stream management: processing high-speed data streams}.
\newblock Springer, 2016.

\bibitem{gvisor}
Open-sourcing g{V}isor, a sandboxed container runtime.
\newblock
  \url{https://cloud.google.com/blog/products/gcp/open-sourcing-gvisor-a-sandboxed-container}
  \url{-runtime}.

\bibitem{han2013network}
S.~Han, N.~Egi, A.~Panda, S.~Ratnasamy, G.~Shi, and S.~Shenker.
\newblock Network support for resource disaggregation in next-generation
  datacenters.
\newblock In {\em Proceedings of the Twelfth ACM Workshop on Hot Topics in
  Networks}, page~10. ACM, 2013.

\bibitem{twitterinfra}
M.~Hashemi.
\newblock The infrastructure behind {T}witter: {S}cale.
\newblock
  \url{https://blog.twitter.com/engineering/en_us/topics/infrastructure/2017/the-infrastructure-behind-twitter-scale.html}.

\bibitem{facebook}
K.~{Hazelwood}, S.~{Bird}, D.~{Brooks}, S.~{Chintala}, U.~{Diril},
  D.~{Dzhulgakov}, M.~{Fawzy}, B.~{Jia}, Y.~{Jia}, A.~{Kalro}, J.~{Law},
  K.~{Lee}, J.~{Lu}, P.~{Noordhuis}, M.~{Smelyanskiy}, L.~{Xiong}, and
  X.~{Wang}.
\newblock Applied machine learning at facebook: A datacenter infrastructure
  perspective.
\newblock In {\em 2018 IEEE International Symposium on High Performance
  Computer Architecture (HPCA)}, pages 620--629, Feb 2018.

\bibitem{he2012zeta}
Y.~He, S.~Elnikety, J.~Larus, and C.~Yan.
\newblock Zeta: Scheduling interactive services with partial execution.
\newblock In {\em Proceedings of the Third ACM Symposium on Cloud Computing},
  pages 1--14, 2012.

\bibitem{campbell}
P.~Helland and D.~Campbell.
\newblock Building on quicksand.
\newblock {\em CoRR}, abs/0909.1788, 2009.

\bibitem{serverless-cidr19}
J.~M. Hellerstein, J.~M. Faleiro, J.~Gonzalez, J.~Schleier{-}Smith,
  V.~Sreekanti, A.~Tumanov, and C.~Wu.
\newblock Serverless computing: One step forward, two steps back.
\newblock In {\em {CIDR} 2019, 9th Biennial Conference on Innovative Data
  Systems Research, Asilomar, CA, USA, January 13-16, 2019, Online
  Proceedings}, 2019.

\bibitem{openlambda}
S.~Hendrickson, S.~Sturdevant, T.~Harter, V.~Venkataramani, A.~C.
  Arpaci-Dusseau, and R.~H. Arpaci-Dusseau.
\newblock Serverless computation with {O}pen{L}ambda.
\newblock In {\em 8th {USENIX} Workshop on Hot Topics in Cloud Computing
  (HotCloud 16)}, Denver, CO, 2016. {USENIX} Association.

\bibitem{hendrickson2016serverless}
S.~Hendrickson, S.~Sturdevant, T.~Harter, V.~Venkataramani, A.~C.
  Arpaci-Dusseau, and R.~H. Arpaci-Dusseau.
\newblock Serverless computation with {O}pen{L}ambda.
\newblock {\em Elastic}, 60:80, 2016.

\bibitem{holt2015claret}
B.~Holt, I.~Zhang, D.~Ports, M.~Oskin, and L.~Ceze.
\newblock Claret: Using data types for highly concurrent distributed
  transactions.
\newblock In {\em Proceedings of the First Workshop on Principles and Practice
  of Consistency for Distributed Data}, page~4. ACM, 2015.

\bibitem{mobilenet}
A.~G. Howard, M.~Zhu, B.~Chen, D.~Kalenichenko, W.~Wang, T.~Weyand,
  M.~Andreetto, and H.~Adam.
\newblock Mobilenets: Efficient convolutional neural networks for mobile vision
  applications.
\newblock {\em CoRR}, abs/1704.04861, 2017.

\bibitem{isard2007dryad}
M.~Isard, M.~Budiu, Y.~Yu, A.~Birrell, and D.~Fetterly.
\newblock Dryad: Distributed data-parallel programs from sequential building
  blocks.
\newblock In {\em Proceedings of the 2Nd ACM SIGOPS/EuroSys European Conference
  on Computer Systems 2007}, EuroSys '07, pages 59--72, New York, NY, USA,
  2007. ACM.

\bibitem{bv-serverless}
E.~Jonas, J.~Schleier-Smith, V.~Sreekanti, C.-C. Tsai, A.~Khandelwal, Q.~Pu,
  V.~Shankar, J.~Menezes~Carreira, K.~Krauth, N.~Yadwadkar, J.~Gonzalez, R.~A.
  Popa, I.~Stoica, and D.~A. Patterson.
\newblock Cloud programming simplified: A {B}erkeley view on serverless
  computing.
\newblock Technical Report UCB/EECS-2019-3, EECS Department, University of
  California, Berkeley, Feb 2019.

\bibitem{pywren}
E.~Jonas, S.~Venkataraman, I.~Stoica, and B.~Recht.
\newblock Occupy the cloud: Distributed computing for the 99{\%}.
\newblock {\em CoRR}, abs/1702.04024, 2017.

\bibitem{ds2}
V.~Kalavri, J.~Liagouris, M.~Hoffmann, D.~Dimitrova, M.~Forshaw, and T.~Roscoe.
\newblock Three steps is all you need: fast, accurate, automatic scaling
  decisions for distributed streaming dataflows.
\newblock In {\em 13th $\{$USENIX$\}$ Symposium on Operating Systems Design and
  Implementation ($\{$OSDI$\}$ 18)}, pages 783--798, 2018.

\bibitem{kempe2003gossip}
D.~Kempe, A.~Dobra, and J.~Gehrke.
\newblock Gossip-based computation of aggregate information.
\newblock In {\em 44th Annual IEEE Symposium on Foundations of Computer
  Science, 2003. Proceedings.}, pages 482--491. IEEE, 2003.

\bibitem{pocket}
A.~Klimovic, Y.~Wang, P.~Stuedi, A.~Trivedi, J.~Pfefferle, and C.~Kozyrakis.
\newblock Pocket: Elastic ephemeral storage for serverless analytics.
\newblock In {\em 13th $\{$USENIX$\}$ Symposium on Operating Systems Design and
  Implementation ($\{$OSDI$\}$ 18)}, pages 427--444, 2018.

\bibitem{kubeless}
Kubeless.
\newblock \url{http://kubeless.io}.

\bibitem{kubernetes}
Kubernetes: Production-grade container orchestration.
\newblock \url{http://kubernetes.io}.

\bibitem{Lamport:1978:TCO:359545.359563}
L.~Lamport.
\newblock Time, clocks, and the ordering of events in a distributed system.
\newblock {\em Commun. ACM}, 21(7):558--565, July 1978.

\bibitem{lamport2001paxos}
L.~Lamport et~al.
\newblock Paxos made simple.
\newblock {\em ACM Sigact News}, 32(4):18--25, 2001.

\bibitem{pretzel}
Y.~Lee, A.~Scolari, B.-G. Chun, M.~D. Santambrogio, M.~Weimer, and
  M.~Interlandi.
\newblock {PRETZEL}: Opening the black box of machine learning prediction
  serving systems.
\newblock In {\em 13th {USENIX} Symposium on Operating Systems Design and
  Implementation ({OSDI} 18)}, pages 611--626, Carlsbad, CA, 2018. {USENIX}
  Association.

\bibitem{meltdown}
M.~Lipp, M.~Schwarz, D.~Gruss, T.~Prescher, W.~Haas, A.~Fogh, J.~Horn,
  S.~Mangard, P.~Kocher, D.~Genkin, Y.~Yarom, and M.~Hamburg.
\newblock Meltdown: Reading kernel memory from user space.
\newblock In {\em 27th {USENIX} Security Symposium ({USENIX} Security 18)},
  2018.

\bibitem{COPS}
W.~Lloyd, M.~Freedman, M.~Kaminsky, and D.~G.~Andersen.
\newblock Don't settle for eventual: Scalable causal consistency for wide-area
  storage with cops.
\newblock In {\em SOSP'11 - Proceedings of the 23rd ACM Symposium on Operating
  Systems Principles}, pages 401--416, 10 2011.

\bibitem{lloyd2013stronger}
W.~Lloyd, M.~J. Freedman, M.~Kaminsky, and D.~G. Andersen.
\newblock Stronger semantics for low-latency geo-replicated storage.
\newblock In {\em Presented as part of the 10th $\{$USENIX$\}$ Symposium on
  Networked Systems Design and Implementation ($\{$NSDI$\}$ 13)}, pages
  313--328, 2013.

\bibitem{mahajan11cacTR}
P.~Mahajan, L.~Alvisi, and M.~Dahlin.
\newblock Consistency, availability, convergence.
\newblock Technical Report TR-11-22, Computer Science Department, University of
  Texas at Austin, May 2011.

\bibitem{light-vm}
F.~Manco, C.~Lupu, F.~Schmidt, J.~Mendes, S.~Kuenzer, S.~Sati, K.~Yasukata,
  C.~Raiciu, and F.~Huici.
\newblock My vm is lighter (and safer) than your container.
\newblock In {\em Proceedings of the 26th Symposium on Operating Systems
  Principles}, pages 218--233. ACM, 2017.

\bibitem{mcgrath2017serverless}
G.~McGrath and P.~R. Brenner.
\newblock Serverless computing: Design, implementation, and performance.
\newblock In {\em 2017 IEEE 37th International Conference on Distributed
  Computing Systems Workshops (ICDCSW)}, pages 405--410. IEEE, 2017.

\bibitem{akbar2017icant}
S.~A. Mehdi, C.~Littley, N.~Crooks, L.~Alvisi, N.~Bronson, and W.~Lloyd.
\newblock I can{\textquoteright}t believe it{\textquoteright}s not causal!
  scalable causal consistency with no slowdown cascades.
\newblock In {\em 14th {USENIX} Symposium on Networked Systems Design and
  Implementation ({NSDI} 17)}, pages 453--468, Boston, MA, Mar. 2017. {USENIX}
  Association.

\bibitem{social-network-zipf}
A.~Mislove, M.~Marcon, K.~P. Gummadi, P.~Druschel, and B.~Bhattacharjee.
\newblock Measurement and analysis of online social networks.
\newblock In {\em Proceedings of the 7th ACM SIGCOMM Conference on Internet
  Measurement}, IMC '07, pages 29--42, New York, NY, USA, 2007. ACM.

\bibitem{sock-atc18}
E.~Oakes, L.~Yang, D.~Zhou, K.~Houck, T.~Harter, A.~Arpaci-Dusseau, and
  R.~Arpaci-Dusseau.
\newblock {SOCK}: Rapid task provisioning with serverless-optimized containers.
\newblock In {\em 2018 {USENIX} Annual Technical Conference ({USENIX} {ATC}
  18)}, pages 57--70, Boston, MA, 2018. {USENIX} Association.

\bibitem{openfaas}
Home | openfaas - serverless functions made simple.
\newblock \url{https://www.openfaas.com}.

\bibitem{openwhisk}
Apache openwhisk is a serverless, open source cloud platform.
\newblock \url{https://openwhisk.apache.org}.

\bibitem{pang2019zanzibar}
R.~Pang, R.~Caceres, M.~Burrows, Z.~Chen, P.~Dave, N.~Germer, A.~Golynski,
  K.~Graney, N.~Kang, L.~Kissner, J.~L. Korn, A.~Parmar, C.~D. Richards, and
  M.~Wang.
\newblock Zanzibar: Google’s consistent, global authorization system.
\newblock In {\em 2019 {USENIX} Annual Technical Conference ({USENIX} {ATC}
  '19)}, Renton, WA, 2019.

\bibitem{perron2019starling}
M.~Perron, R.~C. Fernandez, D.~DeWitt, and S.~Madden.
\newblock Starling: A scalable query engine on cloud function services.
\newblock {\em arXiv preprint arXiv:1911.11727}, 2019.

\bibitem{pu2019shuffling}
Q.~Pu, S.~Venkataraman, and I.~Stoica.
\newblock Shuffling, fast and slow: Scalable analytics on serverless
  infrastructure.
\newblock In {\em 16th $\{$USENIX$\}$ Symposium on Networked Systems Design and
  Implementation ($\{$NSDI$\}$ 19)}, pages 193--206, 2019.

\bibitem{Ratnasamy:2001:SCN:964723.383072}
S.~Ratnasamy, P.~Francis, M.~Handley, R.~Karp, and S.~Shenker.
\newblock A scalable content-addressable network.
\newblock {\em SIGCOMM Comput. Commun. Rev.}, 31(4):161--172, Aug. 2001.

\bibitem{Raynal:1996:LTC:226705.226711}
M.~Raynal and M.~Singhal.
\newblock Logical time: Capturing causality in distributed systems.
\newblock {\em Computer}, 29(2):49--56, Feb. 1996.

\bibitem{retwis}
Tutorial: Design and implementation of a simple {T}witter clone using php and
  the redis key-value store | redis.
\newblock \url{https://redis.io/topics/twitter-clone}.

\bibitem{retwis-py}
pims/retwis-py: Retwis clone in python.
\newblock \url{https://github.com/pims/retwis-py}.

\bibitem{rodruigues2003one}
R.~Rodruigues, A.~Gupta, and B.~Liskov.
\newblock One-hop lookups for peer-to-peer overlays.
\newblock In {\em Proceedings of the 11th Workshop on Hot Topics in Operating
  Systems (HotOS’03)}, 2003.

\bibitem{rowstron2001pastry}
A.~Rowstron and P.~Druschel.
\newblock Pastry: Scalable, decentralized object location, and routing for
  large-scale peer-to-peer systems.
\newblock In {\em IFIP/ACM International Conference on Distributed Systems
  Platforms and Open Distributed Processing}, pages 329--350. Springer, 2001.

\bibitem{sandserverless}
{Nokia Bell Labs Project SAND}.
\newblock \url{https://sandserverless.org}.

\bibitem{numpywren}
V.~Shankar, K.~Krauth, Q.~Pu, E.~Jonas, S.~Venkataraman, I.~Stoica, B.~Recht,
  and J.~Ragan{-}Kelley.
\newblock numpywren: serverless linear algebra.
\newblock {\em CoRR}, abs/1810.09679, 2018.

\bibitem{shapiro2011conflict}
M.~Shapiro, N.~Pregui{\c{c}}a, C.~Baquero, and M.~Zawirski.
\newblock Conflict-free replicated data types.
\newblock In {\em Symposium on Self-Stabilizing Systems}, pages 386--400.
  Springer, 2011.

\bibitem{archipelago}
A.~Singhvi, K.~Houck, A.~Balasubramanian, M.~D. Shaikh, S.~Venkataraman, and
  A.~Akella.
\newblock Archipelago: {A} scalable low-latency serverless platform.
\newblock {\em CoRR}, abs/1911.09849, 2019.

\bibitem{sovran2011transactional}
Y.~Sovran, R.~Power, M.~K. Aguilera, and J.~Li.
\newblock Transactional storage for geo-replicated systems.
\newblock In {\em Proceedings of the Twenty-Third ACM Symposium on Operating
  Systems Principles}, pages 385--400. ACM, 2011.

\bibitem{stoica2001chord}
I.~Stoica, R.~Morris, D.~Karger, M.~F. Kaashoek, and H.~Balakrishnan.
\newblock Chord: A scalable peer-to-peer lookup service for internet
  applications.
\newblock {\em ACM SIGCOMM Computer Communication Review}, 31(4):149--160,
  2001.

\bibitem{taft-thesis}
R.~R.~Y. Taft.
\newblock {\em Elastic database systems}.
\newblock PhD thesis, Massachusetts Institute of Technology, 2017.

\bibitem{terry1994session}
D.~B. Terry, A.~J. Demers, K.~Petersen, M.~J. Spreitzer, M.~M. Theimer, and
  B.~B. Welch.
\newblock Session guarantees for weakly consistent replicated data.
\newblock In {\em Proceedings of 3rd International Conference on Parallel and
  Distributed Information Systems}, pages 140--149. IEEE, 1994.

\bibitem{van2017spec}
E.~Van~Eyk, A.~Iosup, S.~Seif, and M.~Th{\"o}mmes.
\newblock The {SPEC} cloud group's research vision on {FaaS} and serverless
  architectures.
\newblock In {\em Proceedings of the 2nd International Workshop on Serverless
  Computing}, pages 1--4. ACM, 2017.

\bibitem{vogels2009eventually}
W.~Vogels.
\newblock Eventually consistent.
\newblock {\em Communications of the ACM}, 52(1):40--44, 2009.

\bibitem{wagnercompute}
T.~A. Wagner.
\newblock Acquisition and maintenance of compute capacity, Sept.~4 2018.
\newblock US Patent 10067801B1.

\bibitem{wang2018peeking}
L.~Wang, M.~Li, Y.~Zhang, T.~Ristenpart, and M.~Swift.
\newblock Peeking behind the curtains of serverless platforms.
\newblock In {\em 2018 $\{$USENIX$\}$ Annual Technical Conference
  ($\{$USENIX$\}$$\{$ATC$\}$ 18)}, pages 133--146, 2018.

\bibitem{wu2019anna}
C.~Wu, J.~Faleiro, Y.~Lin, and J.~Hellerstein.
\newblock Anna: A kvs for any scale.
\newblock {\em IEEE Transactions on Knowledge and Data Engineering}, 2019.

\bibitem{anna-vldb}
C.~Wu, V.~Sreekanti, and J.~M. Hellerstein.
\newblock Autoscaling tiered cloud storage in anna.
\newblock {\em PVLDB}, 12(6):624--638, 2019.

\bibitem{yan2018carousel}
X.~Yan, L.~Yang, H.~Zhang, X.~C. Lin, B.~Wong, K.~Salem, and T.~Brecht.
\newblock Carousel: low-latency transaction processing for globally-distributed
  data.
\newblock In {\em Proceedings of the 2018 International Conference on
  Management of Data}, pages 231--243. ACM, 2018.

\bibitem{zaharia2012resilient}
M.~Zaharia, M.~Chowdhury, T.~Das, A.~Dave, J.~Ma, M.~McCauley, M.~J. Franklin,
  S.~Shenker, and I.~Stoica.
\newblock Resilient distributed datasets: A fault-tolerant abstraction for
  in-memory cluster computing.
\newblock In {\em Proceedings of the 9th USENIX conference on Networked Systems
  Design and Implementation}, pages 2--2. USENIX Association, 2012.

\bibitem{zamanian2016end}
E.~Zamanian, C.~Binnig, T.~Kraska, and T.~Harris.
\newblock The end of a myth: Distributed transactions can scale.
\newblock {\em arXiv preprint arXiv:1607.00655}, 2016.

\bibitem{Zawirski:2015:WFR:2814576.2814733}
M.~Zawirski, N.~Pregui\c{c}a, S.~Duarte, A.~Bieniusa, V.~Balegas, and
  M.~Shapiro.
\newblock Write fast, read in the past: Causal consistency for client-side
  applications.
\newblock In {\em Proceedings of the 16th Annual Middleware Conference},
  Middleware '15, pages 75--87, New York, NY, USA, 2015. ACM.

\bibitem{zhang2016diamond}
I.~Zhang, N.~Lebeck, P.~Fonseca, B.~Holt, R.~Cheng, A.~Norberg,
  A.~Krishnamurthy, and H.~M. Levy.
\newblock Diamond: Automating data management and storage for wide-area,
  reactive applications.
\newblock In {\em 12th $\{$USENIX$\}$ Symposium on Operating Systems Design and
  Implementation ($\{$OSDI$\}$ 16)}, pages 723--738, 2016.

\bibitem{zhang2019narrowing}
T.~Zhang, D.~Xie, F.~Li, and R.~Stutsman.
\newblock Narrowing the gap between serverless and its state with storage
  functions.
\newblock In {\em Proceedings of the ACM Symposium on Cloud Computing}, pages
  1--12, 2019.

\end{thebibliography}
